\documentclass[3p,times,twocolumn]{elsarticle}

\usepackage{amssymb}
\usepackage{amsthm}
\usepackage{amsmath}

\usepackage[figuresright]{rotating}

\usepackage{subfig}

\usepackage{graphicx}
\usepackage{dcolumn}
\usepackage{bm}
\usepackage{epstopdf}
\usepackage{epsfig}
\usepackage[colorlinks = true,
linkcolor = blue,
urlcolor  = blue,
citecolor = blue,
anchorcolor = blue]{hyperref}
\usepackage{url}

\DeclareMathOperator{\sech}{sech}

\newcommand{\R}{\mathbb{R}}
\newcommand{\C}{\mathbb{C}}
\newcommand{\Z}{\mathbb{Z}}

\newcommand{\eps}{\varepsilon}

\newtheorem{theorem}{Theorem}
\newtheorem{lemma}{Lemma}

\begin{document}
	
	\begin{frontmatter}
		
		\title{Exponential asymptotics of dark and bright solitons in the discrete nonlinear Schr\"odinger equation}
		
		
		\author[au1]{F.T.\ Adriano}
		\author[au1]{A.N.\ Hasmi}
		\author[au2,au2b]{R.\ Kusdiantara}
		\author[au1,au3]{H.\ Susanto\corref{cor1}}
		\ead{hadi.susanto@yandex.com}
		
		\address[au1]{Department of Mathematics, Khalifa University, PO Box 127788, Abu Dhabi, United Arab Emirates}
		\cortext[cor1]{Corresponding author}
		\address[au2]{Industrial and Financial Mathematics Research Group, Institut Teknologi Bandung,\\ Jl.\ Ganesha No.\ 10, Bandung, 40132, Indonesia}
		\address[au2b]{Centre of Mathematical Modelling and Simulation, Institut Teknologi Bandung,\\ Jl.\ Ganesha No.\ 10, Bandung, 40132, Indonesia}
		\address[au3]{Department of Mathematics, Faculty of Mathematics and Natural Sciences, Universitas Indonesia,\\ Gedung D Lt.\ 2 FMIPA Kampus UI Depok, 16424, Indonesia}

		\begin{abstract}
			We investigate the existence and linear stability of solitons in the nonlinear Schr\"odinger lattices in the strong coupling regime. Focusing and defocusing nonlinearities are considered, giving rise to bright and dark solitons. In this regime, the effects of lattice discreteness become exponentially small, requiring a beyond-all-orders analysis. To this end, we employ exponential asymptotics to derive soliton solutions and examine their stability systematically. We show that only two symmetry-related soliton configurations are permissible: onsite solitons centered at lattice sites and intersite solitons positioned between adjacent sites. {Although the instability of intersite solitons due to real eigenvalue pairs is known numerically, a rigorous analytical account, \emph{particularly for dark solitons}, has been lacking. Our work fills this gap, yielding analytical predictions that match numerical computations with high accuracy. We also establish the linear stability of onsite bright solitons. While the method cannot directly resolve the quartet eigenvalue-induced instability of onsite dark solitons due to the continuous spectrum covering the entire imaginary axis, we conjecture an eigenvalue-counting argument that supports their instability}. Overall, our application of the exponential asymptotics method shows the versatility of this approach for addressing multiscale problems in discrete nonlinear systems.
		\end{abstract}
		
		
	\end{frontmatter}
	\pagebreak
	\section{Introduction}
	
	Nonlinear waves on discrete systems play a crucial role in various physical contexts, including nonlinear optics, condensed matter physics, and biophysics \cite{kivshar2003optical,braun2004frenkel,kevrekidis2009discrete,kevrekidis2015defocusing}. 
	These phenomena are often modeled by differential-difference equations (DDEs), where systems evolve continuously in time and discretely in space, akin to coupled differential equations on a lattice. 
	A key example in this category is the discrete nonlinear Schr\"odinger equation (DNLS), which describes the evolution of wave fields in discrete media, such as arrays of coupled nonlinear waveguides or Bose-Einstein condensates:
	\begin{equation} \label{eqn:dnls_main_cubic_focusing_defocusing}
		i\partial_{t}\phi = -\dfrac{C}{2}\Delta \phi \mp \mu \phi \pm |\phi|^{2}\phi,
	\end{equation}
	where $t \in \R$ and $\phi = \phi(n) \in \C$ for $n \in \Z$. Here, $C$ denotes the coupling strength between lattice sites, and $\Delta$ is the discrete Laplacian operator:
	\begin{align*}
		\Delta \phi(n) &= \phi(n+1) - 2\phi(n) + \phi(n-1).
	\end{align*}
	
	The sign of the nonlinearity determines its type, focusing on the negative sign and defocusing on the positive sign. Focusing nonlinearity gives rise to bright solitons, while defocusing nonlinearity leads to dark solitons/kinks. These solitons represent localized, self-trapped structures in discrete media and have been extensively used to model discrete soliton propagation \cite{Aceves_94_dnls_modelling}.
	
	Studies on the existence of steady-state solutions of the {focusing} DNLS and their stability have been done extensively (see, e.g., \cite{kevrekidis2009discrete,kevrekidis2015defocusing}). In the anticontinuum limit ($C \to 0$), it was shown by Weinstein \cite{Weinstein1999} using the fact that the DNLS is a Hamiltonian system with a time-independent minimizer called one-site breather. Kapitula and Kevrekidis \cite{Kapitula2001} studied the structure of a minimizer in the continuum limit ($C \to \infty$) and its linear stability. In the continuum limit, one could approximate the DNLS as a perturbation of the NLS. However, this approximation fails to be valid since this implies that a translationally invariant solution exists for the DNLS. However, it is known that discretization for the DNLS \eqref{eqn:dnls_main_cubic_focusing_defocusing} breaks the translational invariant nature of the NLS, as such, there exist two types of steady-state solutions, i.e., the site-centered (onsite) and the mode-centered (off-site or intersite) solutions \cite{Kapitula2001}. Ultimately, any order of perturbative approximation of the focusing DNLS from the NLS would fail to capture this phenomenon since this would only account for the effects of algebraic order (in $1/\sqrt{C}$). In contrast, the saddle-node bifurcation, which causes the birth of the two types of solutions, requires calculations of exponentially small order as shown by Fiedler and Scheurle \cite{fiedler1996discretization}. Therefore, to fully understand the behavior of the solutions of the DNLS \eqref{eqn:dnls_main_cubic_focusing_defocusing}, we have to deal with the original discrete equation and take into account the exponentially small discretization effects.
	
	In the case of bright solitons, Kapitula and Kevrekidis \cite{Kapitula2001} dealt with the finer details of discreteness by employing a Melnikov function approach to the problem. The focusing DNLS \eqref{eqn:dnls_main_cubic_focusing_defocusing} is regarded as a perturbation of an integrable discretization of the NLS, namely the Ablowitz-Ladik DNLS (AL-DNLS) \cite{Ablowitz1975}. They constructed the appropriate Melnikov function from the integrable equation to determine the existence of bright soliton solutions. They showed that exponentially small terms appear upon evaluation of the Melnikov sums. The vanishing of these sums corresponds to the two solutions of the DNLS, i.e., the onsite and off-site bright solitons. 
	
	The discrete nature of the DNLS also poses a problem when determining the linear stability of the solutions. The eigenvalues of the linearized operator for the two steady-state solutions are also exponentially small. Kapitula and Kevrekidis \cite{Kapitula2001} showed this in the case of bright solitons. They developed a general method using the Evans function for nonautonomous linear difference equations. They utilized it to calculate the linearized operator's spectrum in the DNLS case. It was shown that the off-site solution is unstable with a pair of real eigenvalues. The onsite solution is stable with a pair of imaginary eigenvalues with an exponentially small magnitude. The stability of the onsite solution was obtained after invoking the result of Grillakis et al.\ \cite{Grillakis1987}. While the developed scheme allows for a thorough analysis of solutions to a DDE and its linear stability, the procedure requires the DDE to have an appropriate non-generic counterpart, one in which the stable and unstable manifolds do not intersect transversely. Then, the DDE may be regarded as a perturbation of the corresponding non-generic DDE.
	
	In contrast to bright solitons, the onsite dark soliton is also unstable for coupling beyond a critical value. This is due to a Hamiltonian Hopf bifurcation, which arises as the coupling strength $C$ passes the critical value, causing oscillatory instability for the onsite solution. The bifurcation was studied through an analysis in the anticontinuum limit (see \cite{PhysRevLett.82.85,PhysRevE.75.066608,Johansson2004,pelinovsky2008stability,susanto2005discrete,yoshimura2019existence} for details). It was also numerically studied by Johansson and Kivshar \cite{PhysRevLett.82.85} that the oscillatory instability persists for larger values of the coupling strength $C$. However, {asymptotic analysis of the dark solitons in the continuum limit is rather completely lacking}.
	
	{The novelty of the present work lies in the application of {exponential asymptotics}, or asymptotics beyond all orders, to the DNLS equation—specifically to analyze the existence and stability of onsite and intersite bright and dark solitons in the focusing and defocusing regimes, respectively. Unlike standard asymptotic or perturbative methods, which are limited to capturing algebraic-order behavior, exponential asymptotics allow us to systematically resolve features that are exponentially small in the perturbation parameter and that play a critical role in phenomena such as wave selection. We employ the method developed by King and Chapman \cite{KING_CHAPMAN_2001} and demonstrate that an exponentially small remainder term, i.e., the {Stokes phenomenon}, appears in the solution to the DNLS equation, except in the case of onsite or off-site solutions. To the best of our knowledge, this is also the first study to apply these techniques to the stability analysis of DNLS discrete solitons, inspired by the work of Hwang et al.\ \cite{HWANG20111055} on the NLS equation with a linear potential. We derive the exponentially small eigenvalues of the corresponding linearized operator by tracking the emergence of growing tails associated with the underlying Stokes structure.}
	
	{For dark solitons, we show, for the first time, to the best of our knowledge, that only two types of localized solutions are possible: onsite and off-site solitons. Our exponential asymptotic analysis reveals the exponential instability of the off-site solitons. However, this method cannot be applied to establish the oscillatory instability of onsite solitons, as the continuous spectrum of the linearized operator spans the entire imaginary axis. Nevertheless, we provide an argument based on the eigenvalue counting theorem that suggests how the instability of onsite solitons may be inferred. We also complement our analytical results with careful numerical computations, which show excellent agreement with the asymptotic predictions.}
	
	The paper is organized as follows. In Section \ref{sec2}, we begin our work with developing the asymptotics beyond all orders method for dark solitons, starting with constructing both onsite and intersite solitons in the strong coupling limit. This is followed by an analysis of the linear stability of these solitons using the same asymptotic approach. Section \ref{sec3} extends the asymptotic framework to bright solitons, applying similar techniques to understand their structure and stability. Section \ref{sec4} discusses the numerical methods we used to obtain computational results. We also provide a detailed comparison with the theoretical predictions to validate the accuracy of the asymptotic expansions. Finally, Section \ref{sec:conclusion} presents the conclusions, summarizing our findings and discussing potential directions for future research.

	\section{Dark Solitons}	\label{sec2}
	We will first consider the case with defocusing nonlinearity, i.e
	\begin{equation} \label{eqn:dnls_main_cubic_defocusing}
		i\partial_{t}\phi = -\dfrac{C}{2}\Delta \phi -\mu \phi + |\phi|^{2}\phi.
	\end{equation}
	By an appropriate scaling, we will assume that $\mu = 1$. We are interested in the steady-state dark solitons of the governing equation. Let us denote them as $\tilde{\phi}$. After that, we will analyze their linear stability.

    To do so, we start by linearizing equation \eqref{eqn:dnls_main_cubic_defocusing} about the stationary solution $\tilde{\phi}$. We write $\phi = \tilde{\phi} + (p+iq)$ where $p(n,t),q(n,t)$ are such that $|p|, |q| \ll 1$, from which we obtain the linearized equation for $p$ and $q$ as
	\begin{align}\label{eqn:dnls_defocusing_linearized_eqn}
		\begin{bmatrix}
			p_{t} \\ q_{t}
		\end{bmatrix}=
		\mathcal{L}		\begin{bmatrix}
			p \\ q
		\end{bmatrix},
	\end{align}
	where 
	\begin{align*}
		\mathcal{L} &= 		\begin{bmatrix}
			O & L_{-} \\
			-L_{+} & O
		\end{bmatrix},\\
		L_{-} &= -\dfrac{C}2
        \Delta - 1 + |\tilde{\phi}|^{2}, \quad 
		L_{+} = -\dfrac{C}2
        \Delta - 1 + 3|\tilde{\phi}|^{2}.
	\end{align*}
	With $p = u e^{\lambda t}$ and $q = v e^{\lambda t}$, we obtain the eigenvalue equation
	\begin{align}\label{eqn:dnls_defocusing_evp}
		\lambda \begin{bmatrix}
			u \\ v
		\end{bmatrix} = \begin{bmatrix}
			O & L_{-} \\ -L_{+} & O
		\end{bmatrix}\begin{bmatrix}
			u \\ v
		\end{bmatrix} = \mathcal{L} \begin{bmatrix}
			u \\ v
		\end{bmatrix}.
	\end{align}

Our main result in this section is stated in the following theorem:
\begin{theorem}\label{thm:darksoliton}
The defocusing DNLS equation \eqref{eqn:dnls_main_cubic_defocusing} in the strong coupling limit $C\to\infty$ admits two types of dark solitons of the form
\[
    \phi(x) \sim \tanh\left(\frac{x}{\sqrt{2}}\right)+ C_{R} \textnormal{Im}\ G(x)\cos(2\pi n_{0}),
\]
where $x = \varepsilon(n - n_{0})$, $\varepsilon=\sqrt{2/C}$, $C_{R} = 2\pi|\Lambda_{1}|\varepsilon^{-4}e^{-\sqrt{2}\pi^{2}/\varepsilon}$ with $\Lambda_{1}$ being given by Eqs.\ \eqref{eqn:recur_relation_Aj} and \eqref{eqn:Lambda1_Aj}, $G(x)$ is defined in Eq.\ \eqref{eqn:f0_G}, and $n_{0}$ can take the values $0$ or $1/2$ (mod $1$). The first value, $n_0 = 0$, corresponds to the onsite soliton, while the second value, $n_0 = 1/2$, corresponds to the off-site (or intersite) soliton. The off-site soliton is exponentially unstable due to a pair of real-valued linear eigenvalues given asymptotically in Eq.\ \eqref{eqn:dnls_dark_ev_approximation}, whereas the onsite soliton is unstable due to an oscillatory instability (see Theorem \ref{thm:eigval_count}).
\end{theorem}

\subsection{Time-independent solutions}

We first analyze the steady-state solution to \eqref{eqn:dnls_main_cubic_defocusing} near the continuum limit $C \to \infty$) or $\varepsilon \to 0$ and consider the time-independent equation of \eqref{eqn:dnls_main_cubic_defocusing} as
	\begin{align} \label{eqn:dnls_main_cubic_defocusing_stationary}
		0 &= \Delta \phi - \varepsilon^{2}(-\phi + |\phi|^{2}\phi).
	\end{align}
	
	We introduce the slow scale $x = \varepsilon(n-n_{0})$ where $n_{0} \in \Z$ is a constant that represents the center of the solution. This constant will be determined a posteriori following the asymptotic analysis.
	Using the slow scale, we may now regard the Laplacian $\Delta$ term to be
	\begin{align*}
		\Delta \phi(x) &= \phi(x + \varepsilon) + \phi(x-\varepsilon) - 2\phi(x) \\
		&= 2\sum_{m \geq 1}\dfrac{\varepsilon^{2m}}{(2m)!}\partial_{x}^{2m}\phi.
	\end{align*}
	Seeing that the small terms occur in powers of $\varepsilon^{2}$, naturally, we consider the expansion for $\phi$ in powers of $\varepsilon^{2}$ as
	\begin{align} \label{eqn:phi_expansion}
		\phi(x) &= \sum_{j=0}^{\infty}\varepsilon^{2j}\phi_{j}(x).
	\end{align}
	Thus, we get at leading order $O(\varepsilon^{2})$, the equation for $\phi_{0}$ as
	\begin{align} \label{eqn:phi0_eqn}
		0 &= \phi_{0}'' - (-\phi_{0} + \phi_{0}^{3}),
	\end{align}
	which gives the leading order kink solution
	\begin{align} \label{eqn:phi0_soln}
		\phi_{0}(x) &= \tanh\left(\frac{x}{\sqrt{2}}\right).
	\end{align}
	At order $O(\varepsilon^{2j+2})$, we have 
	\begin{align} \label{eqn:phij_eqn}
		0 =\ &\sum_{p=1}^{j+1}\dfrac{1}{(2p)!}\partial_{x}^{2p}\phi_{j-p+1} + \phi_{j} + \sum_{k=0}^{j}\sum_{l=0}^{k}\phi_{l}\phi_{k-l}\phi_{j-k}.
	\end{align}
	Even though, in this case, $x \in \R$, we continue the solution to the complex plane, i.e., $x \in \C$, and we make the following important observation. On the complex plane, $\phi_{0}(x)$ has poles of order 1 at
	\begin{align} \label{eqn:dnls_defocusing_leading_order_poles}
		\zeta_{k} &= i \dfrac{1}{\sqrt{2}}(2k+1)\pi \quad (k \in \Z).
	\end{align}
	Near these poles, $\phi_{0}(x)$ diverges as
	\begin{align} \label{eqn:dnls_defocusing_leading_order_behaviour_near_poles}
		\phi_{0}(x) \sim \dfrac{\sqrt{2}}{x-\zeta_{k}}.
	\end{align}
	It now follows from \eqref{eqn:phij_eqn} by considering successive equations for $j = 1,2,\dots,$ that $\phi_{j}$ inherits these singularities at the same set of poles. Moreover, as $\phi_{j}$ is determined partly by differentiating $\phi_{j-1}$ 4 times, $\phi_{j-2}$ 6 times, and so on, and then integrating twice, we see that the order of the poles would increase by two, that is if $\phi_{j-1}$ has poles of order $l$, then $\phi_{j}$ would have poles of order $l+2$. Therefore, $\phi_{j}$ would have poles of order $2j+1$. This process of repeated differentiation also means that each successive $j$ would introduce a factorial term, leading to $\phi_{j}$ being divergent as $j$ increases by taking the general form of a factorial over power.
	
	We shall see that the diverging behavior of the $\phi_{j}(x)$ terms as $j \to \infty$ in the expansion would cause a Stokes phenomenon to occur, where exponentially small remainder terms are switched on as the corresponding Stokes line is crossed. However, as this remainder term is exponentially small, it escapes every term in the expansion \eqref{eqn:phi_expansion}, which is algebraic in $\varepsilon$. This necessitates the need to analyze exponentially small effects, which can be done by the methods of exponential asymptotics developed by King and Chapman \cite{KING_CHAPMAN_2001}.

	\subsection{Late Order Terms}
	We will now analyze the diverging behavior of $\phi_{j}$ as $j \to \infty$. Thus, we will analyze the late-order terms' behavior in the expansion. Following the previous discussion, we expect that $\phi_{j}$ would take on the form of a factorial over power as $j \to \infty$, that is 
	\begin{align} \label{eqn:phij_factorial_power_ansatz}
		\phi_{j}(x) \sim (-1)^{j} \dfrac{\Gamma(2j+\beta)}{[W(x)]^{2j+\beta}}f_{0}(x),
	\end{align} 
	as $j \to \infty$. Here, $\beta$ is a constant to be determined by matching, and $W$ represents the singularity at the poles $\zeta = \zeta_{k}$. Thus, $W$ vanishes at $x = \zeta$. The function $f_0(x)$ is currently unknown and will be specified later (see Eqs.\ \eqref{eqn:f0_eqn}, \eqref{eqn:f0_g}, and \eqref{eqn:f0_G}).
	\par
	At order $O(\varepsilon^{2j+2})$, from \eqref{eqn:phij_eqn}, we have the equation for $\phi_{j}$ as 
	\begin{align} \label{eqn:phij_eqn_j_large}
		0 &= \sum_{p=1}^{j+1}\dfrac{1}{(2p)!}\partial_{x}^{2p}\phi_{j-p+1} - (-1 + 3\phi_{0}^{2})\phi_{j} + \dots,
	\end{align}
	as $j \to \infty$. In light of the factorial over power ansatz \eqref{eqn:phij_factorial_power_ansatz}, the omitted terms on \eqref{eqn:phij_eqn_j_large} are of order $O\left(1/n\right)$ of the remaining terms.
	\par
	Substituting the ansatz \eqref{eqn:phij_factorial_power_ansatz} to \eqref{eqn:phij_eqn_j_large}, at leading order $O(\Gamma(2j+\beta+2))$, we would obtain
	\begin{align*}
		0 = \sum_{p=1}^{j+1}\dfrac{(-1)^{p}(W')^{2p}}{(2p)!} \approx \sum_{p=1}^{\infty}\dfrac{(-1)^{p}(W')^{2p}}{(2p)!} = \cos(W') - 1
	\end{align*}
	as $j \to \infty$.
	This gives $W' = \kappa = \kappa_{M} = 2M\pi$, $M \in \Z$ and thus $W = \kappa(x-\zeta)$. We may take $M > 0$ without loss of generality by modifying the sign of the function $f_{0}$ accordingly.
	\par
	Continuing at higher orders, after division by $\Gamma(2j+\beta+2)$, we have as $j \to \infty$
	\begin{align*}
		0 = \ &\Bigg\{\dfrac{(-1)^{j+2}}{W^{2j+\beta+2}}\left[(\cos\kappa -1)f_{0}\right] \\
		&+ \dfrac{(-1)^{j+1}}{(2j+\beta+1)W^{2j+\beta+1}}(\sin \kappa) f_{0}' \\
		&+ \dfrac{(-1)^{j}}{(2j+\beta)(2j+\beta+1)W^{2j+\beta}} \cos\kappa f_{0}''\Bigg\} \\
		&- \dfrac{(-1)^{j}}{(2j+\beta)(2j+\beta+1)W^{2j+\beta}}(-1 + 3\phi_{0}^{2}) f_{0}.
	\end{align*}
	The $\sin \kappa$ terms vanish automatically since $\kappa = 2M\pi$ from the leading order. We are now left with the equation for $f_{0}$ at $O(1/j^{2})$ as
	\begin{align} \label{eqn:f0_eqn}
		f_{0}'' - (-1 + 3\phi_{0}^{2})f_{0} &= 0,
	\end{align}
	whose solutions are
	\begin{align}
		f_{0}(x) &= \lambda_{M} g(x) = \lambda_{M}\phi_{0}'(x), \label{eqn:f0_g}\\
		f_{0}(x) &= \Lambda_{M}G(x) = \Lambda_{M} \phi_{0}'(x) \int_{\zeta}^{x}[\phi_{0}'(s)]^{-2} ds, \label{eqn:f0_G}
	\end{align}
	where $\lambda_{M},\Lambda_{M}$ are constants which depend on $\kappa = \kappa_{M} = 2M\pi$.
	\par
	We can now determine $\beta$ by matching the order of the poles of $\phi_{j}$. From the discussion about the poles of $\phi_{j}$, we know that $\phi_{j}$ would have a pole of order $2j+1$ at $\zeta_{k}$. On the other hand, from the late order form \eqref{eqn:phij_factorial_power_ansatz}, if $f_{0}(x) = \lambda_{M}g(x)$, we have that the pole of $\phi_{j}$ is of order $2j+\beta + 2$ (since $g(x) = \phi_{0}'(x) \sim (x-\zeta)^{-2}$ as $x \to \zeta$), thus if $f_{0}(x) = \lambda_{M}g(x)$, we have $\beta = -1$. Similarly, we can deduce that if $f_{0}(x) = \Lambda_{M}G(x)$, then the pole of $\phi_{j}$ is of order $2j+\beta-3$, thus in this case $\beta = 4$. Therefore, we have that
	\begin{align*}
		\phi_{j}(x) \sim \sum_{M=1}^{\infty}\Bigg[&\dfrac{(-1)^{j}\Gamma(2j-1)\lambda_{M}g(x)}{[2M\pi(x-\zeta)]^{2j-1}} \\
		&+ \dfrac{(-1)^{j}\Gamma(2j+4)\Lambda_{M}G(x)}{[2M\pi(x-\zeta)]^{2j+4}}\Bigg].
	\end{align*}
	However, we note that the dominant contribution as $j \to \infty$ would come from the second term, with $f_{0}(x) = \Lambda_{M}G(x)$ with $M = 1$, i.e.,
	\begin{align} \label{eqn:phij_dominant}
		\phi_{j}(x) &\sim \dfrac{(-1)^{j}\Gamma(2j+4)\Lambda_{1}G(x)}{[2\pi(x-\zeta)]^{2j+4}},
	\end{align}
	as $j \to \infty$.
	\par
	The constant $\Lambda_{1}$ can now be determined via matching near the pole, $x \to \zeta$. This matching procedure can be done by defining an appropriate inner region near the pole. 
	
	\begin{figure*}[tbhp]
		\centering
		\includegraphics[width=0.7\linewidth]{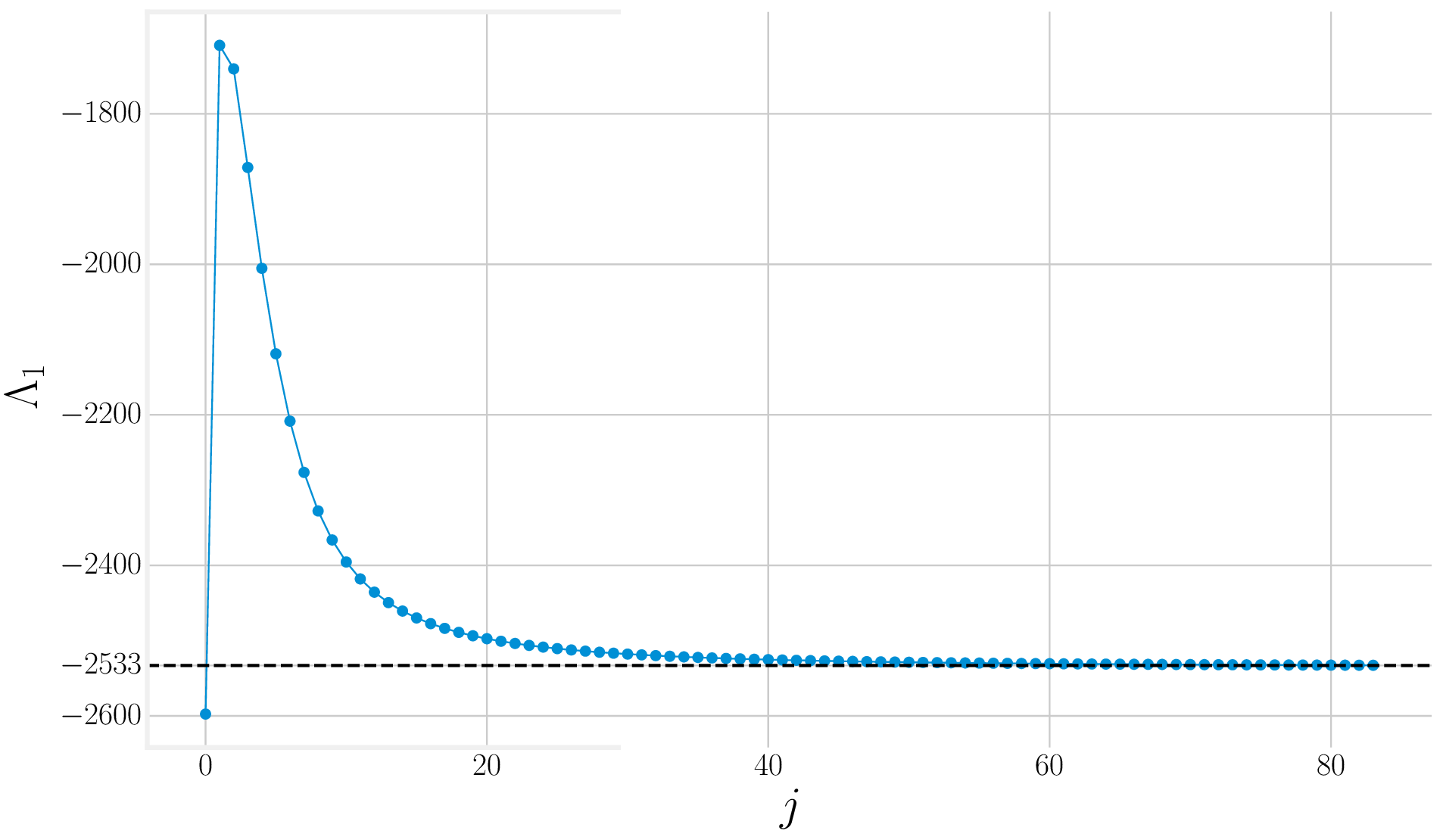}
		\caption{{Numerical calculation of $\Lambda_{1}$, showing its convergence as the number of terms increases. $\Lambda_{1}$ quantifies the exponentially small contributions to the soliton solution.}}
		\label{fig:L1_approx}
	\end{figure*}
	
	We first notice that the expansion \eqref{eqn:phi_expansion} fails to be valid when $x - \zeta = O(\varepsilon)$ since the terms $\varepsilon^{2j}\phi_{j}(x)$ is of the same order as the leading order $\phi_{0}(x)$ as $x - \zeta = O(\varepsilon)$. This and the fact that at leading order $\phi_{0} \sim O((x-\zeta)^{-1})$ as $x \to \zeta$ suggests the following scaling near the pole $\zeta$,
	\begin{align} \label{eqn:inner_variables}
		x = \zeta + \varepsilon q,  \quad\varphi(q) = \varepsilon \phi(x),
	\end{align}
	where $q = O(1)$ is the inner variable. With the inner variable $q$ and the scaled function $\varphi$, we may derive an inner equation by substituting this into the stationary DNLS equation \eqref{eqn:dnls_main_cubic_defocusing_stationary} to give
	\begin{align} \label{eqn:inner_eqn}
		\varphi(q+1) - 2\varphi(q) + \varphi(q-1) - (\varphi^{3}-\varepsilon^{2}\phi) = 0.
	\end{align}
	We also expand $\varphi$ in powers of $\varepsilon^{2}$ as $\varphi = \varphi_{0} + O(\varepsilon^{2})$, which at leading order gives the equation for $\phi_{0}$ as
	\begin{align} \label{eqn:inner_eqn_leading_order}
		\varphi_{0}(q+1)-2\varphi_{0}(q)+\varphi_{0}(q-1) - \varphi_{0}^{3}(q) = 0.
	\end{align}
	Now, we seek an asymptotic solution to \eqref{eqn:inner_eqn_leading_order} of the form
	\begin{align}\label{eqn:varphi0_expansion}
		\varphi_{0}(q) = \sum_{j=0}^{\infty}\dfrac{A_{j}}{q^{2j+1}},
	\end{align}
	where $|q| \to \infty$, to match with the outer expansion \eqref{eqn:phi_expansion}. Matching the expansions \eqref{eqn:varphi0_expansion} with \eqref{eqn:dnls_defocusing_leading_order_behaviour_near_poles} at leading order (order $O(q^{-1})$) gives that $A_{0} = \sqrt{2}$. The coefficients $A_{j}$ are to be matched with the dominant coefficient of $\phi_{j}$ \eqref{eqn:phij_dominant} as $x \to \zeta$, and thus determine $\Lambda_{1}$ by matching the terms for $j \to \infty$. The coefficients $A_{j}$ may now be determined by solving the recurrence relation obtained by substituting the expansion \eqref{eqn:varphi0_expansion} to \eqref{eqn:inner_eqn_leading_order}, which gives at $O(q^{-(2j+3)})$ the recurrence relation
	\begin{equation} \label{eqn:recur_relation_Aj}
		0 = 2\sum_{p=1}^{j+1}\binom{2j+2}{2p}A_{j-p+1} - \sum_{k=0}^{j}\sum_{l=0}^{k}A_{l}A_{k-l}A_{j-k},
	\end{equation}
	with $A_{0} = \sqrt{2}$. 
	\par
	Now, as $x \to \zeta$, the dominant term in $\phi_{j}$ \eqref{eqn:phij_dominant} is
	\begin{align} \label{eqn:phij_dominant_near_pole}
		\phi_{j} \sim \dfrac{\sqrt{2}(-1)^{j+1}\Gamma(2j+4)\Lambda_{1}}{10(2\pi)^{2j+4}(x-\zeta)^{2j+1}},
	\end{align}
	as $j \to \infty$. Matching this with $A_{j}$ gives
	\begin{align} \label{eqn:Lambda1_Aj}
		\Lambda_{1} = \dfrac{10}{\sqrt{2}}\lim_{j \to \infty} \dfrac{(-1)^{j+1}(2\pi)^{2j+4}}{\Gamma(2j+4)}A_{j}.
	\end{align}
	Solving the recurrence relation numerically, it is found that $\Lambda_{1} \approx -2533$, see Fig.\ \ref{fig:L1_approx}.
	
	\subsection{Remainder Term}
	To determine the exponentially small remainder term, which is switched on by the Stokes phenomenon as we cross the Stokes line passing through the poles of $\phi_{j}(x)$, we truncate the expansion for $\phi(x)$ optimally, that is we truncate the expansion \eqref{eqn:phi_expansion} after $N$ terms such that the remainder is minimized  \cite{berry1990hyperasymptotics}. We do the truncation as follows. Letting $R_{N}$ be the remainder of \eqref{eqn:phi_expansion} after taking $N$ terms, we have the truncated expansion:
	\begin{equation} \label{eqn:expansion_remainder}
		\phi = \sum_{j=0}^{N-1}\varepsilon^{2j}\phi_{j}(x) + R_{N},
	\end{equation}
	where $N \gg 1$ is such that
	\begin{align*}
		\left|\dfrac{\varepsilon^{2N+2}\phi_{N+1}}{\varepsilon^{2N}\phi_{N}}\right| \sim 1.
	\end{align*}
	This effectively gives 
	\begin{equation}
		N \sim \dfrac{|\kappa(x-\zeta)|}{2\varepsilon} + \nu,\label{truncate}
	\end{equation}    
	where $\nu = O(1)$ such that $N \in \Z$. Note again that the asymptotic series \eqref{eqn:phi_expansion} becomes divergent when crossing a Stokes line. In a seminal contribution, Berry \cite{berry1989uniform} introduced the idea of truncating such divergent series at the point where the errors are smallest, known as the optimal (near the least term) truncation. It is clear that the truncation point will depend on the spatial variable $x$, see, e.g., \eqref{truncate}. This procedure leads to the emergence of exponentially small contributions. By discarding the remainder beyond the optimal truncation point, Berry demonstrated that the resulting approximation achieves an exponentially small error. This method, which he referred to as superasymptotics or asymptotics beyond all orders, provides a systematic way to extract meaningful results from divergent expansions (see, e.g., a review \cite{say2016exponential}).

	Substituting the truncated expansion \eqref{eqn:expansion_remainder} to the stationary equation \eqref{eqn:dnls_main_cubic_defocusing_stationary}, we get the equation for the remainder term $R_{N}$ as
	\begin{equation} \label{eqn:remainder_eqn}
		\begin{aligned}
			&\Delta R_{N} - \varepsilon^{2}(-1 + 3\phi_{0}^{2})R_{N} = \\
			&-2\sum_{m=1}^{\infty}\varepsilon^{2(N+m)}\sum_{p=m+1}^{N+m} \dfrac{\partial_{x}^{2p}\phi_{N+m-p}}{(2p)!} + \dots,
		\end{aligned}
	\end{equation}
	where the omitted terms are negligible as $N \gg 1$. It may be checked that the chosen value of $N$ would minimize the right hand side of \eqref{eqn:remainder_eqn}, thus justifying the choice of $N$.
	
	We see that the homogeneous version of \eqref{eqn:remainder_eqn} has the solutions which are asymptotically $R_{N} \sim e^{-i2l\pi x/\varepsilon}f_{0}(x)$ as $\varepsilon \to 0$ where $f_{0}(x) = g(x)$ or $f_{0}(x) = G(x)$. Due to the dominant term in the late order term \eqref{eqn:phij_dominant}, a multiple of the latter will be switched on across the Stokes line with $l = M = 1$.
	\par
	We now use the expression for the late order terms $\phi_{j}$ \eqref{eqn:phij_dominant} to calculate the right-hand side of \eqref{eqn:remainder_eqn}. However for convenience, we still write $\phi_{j}$ in the form of \eqref{eqn:phij_factorial_power_ansatz}, where it is to be understood that $f_{0}(x) = \Lambda_{1} G(x)$ and $W(x) = \kappa(x-\zeta) = 2\pi(x-\zeta)$. After reindexing the outer sum to start from $m = 0$ and invoking Stirling's formula for the Gamma function \cite{abramowitz+stegun}, we have the right-hand side of \eqref{eqn:remainder_eqn} to be
	\begin{align*}
		2 \sum_{m=0}^{\infty}&\varepsilon^{2N+2}\varepsilon^{2m}\sum_{p=m+2}^{N+m+1}\Bigg\{\dfrac{(-1)^{N+m+1+p}}{(2p)!}\times\\
		&\dfrac{\sqrt{2\pi}(2N+\beta+2m+2)^{2N+\beta+2m+3/2}e^{-(2N+\beta+2m+2)}}{W^{2N+\beta+2m+2}}\Bigg\}.
	\end{align*}
	Now, since $2N \gg \beta + 2m + 2$ as $N \to \infty$, we have
	\begin{align*}
		&(2N+\beta+2m+2)^{2N+\beta+2m+2-1/2}e^{-(2N+\beta+2m+2)}\\
		&\sim (2N)^{2N+\beta+2m+2-1/2}e^{-2N}.
	\end{align*}
	We may now pull the terms 
	\begin{align*}
		\dfrac{-2\sqrt{2\pi}(-1)^{N+1}\varepsilon^{2N+2}(2N)^{2N+\beta+3/2}e^{-2N}}{W^{2N+\beta+2}}
	\end{align*}
	out of the summation and calculate the sum as
	\begin{align*}
		&\underbrace{\sum_{m=0}^{\infty}\sum_{p=m+2}^{\infty}}_{\text{Reverse order}} \left[(-1)^{m}\left(\dfrac{\varepsilon(2N)}{W}\right)^{2m}\right]\dfrac{(-1)^{p}\kappa^{2p}}{(2p)!} f_{0} \\
		= \ &\left[\sum_{p=2}^{\infty}\dfrac{(-1)^{p}\kappa^{2p}}{(2p!)}\sum_{m=0}^{p-2}(-1)^{m}\left(\dfrac{\varepsilon(2N)}{W}\right)^{2m}\right]f_{0} \\
		= \ &\dfrac{W^{4}}{(2N\varepsilon)^{2}[W^{2}+(2N\varepsilon)^{2}]} \left(\cosh\left(\dfrac{2\kappa N\varepsilon}{W}\right) - 1\right)f_{0}.
	\end{align*}
	Thus, the right hand side of \eqref{eqn:remainder_eqn} is now
	\begin{align*}
		\dfrac{(-1)^{N}2\sqrt{2\pi}\varepsilon^{2N}(2N)^{2N+\beta-1/2}e^{-2N}}{W^{2N+\beta-2}\left[W^{2}+(2N\varepsilon)^{2}\right]}\left(\cosh\left(\dfrac{2\kappa N\varepsilon}{W}\right) - 1\right)f_{0}.
	\end{align*}
	We introduce the polar coordinate $W = \kappa(x-\zeta) = \rho e^{i\theta}$, thus $N = \frac{\rho}{2\varepsilon} + \nu$. Now, we consider the prefactor to the $\cosh$ term in the coordinates $(\rho,\theta)$. As $\varepsilon \to 0$, we have
	\begin{align*}
		&\dfrac{\varepsilon^{2N}(2N)^{2N+\beta-1/2}e^{-2N}}{W^{2N+\beta-2}[W^{2}+(2N\varepsilon)^{2}]} \\
		&\sim \dfrac{\varepsilon^{2N}(2N)^{2N+\beta-1/2}e^{-2N}}{\rho^{2N+\beta-2}e^{i(2N+\beta-2)\theta}[\rho^{2}e^{i2\theta}+4\varepsilon^{2}N^{2}]} \\
		&\sim \dfrac{\varepsilon^{1/2-\beta}\left(2\varepsilon N/\rho\right)^{2N+\beta-1/2} e^{-i\theta(2N+\beta-2)} e^{-\rho/\varepsilon-2\nu}}{\rho^{1/2}[e^{i2\theta} + 4\varepsilon^{2}N^{2}]} \\
		&\sim \dfrac{\varepsilon^{1/2-\beta}\left(1 + \frac{2\varepsilon\nu}{\rho}\right)^{\rho/\varepsilon + 2\nu + \beta - 1/2} e^{-i\theta(2N+\beta-2)}e^{-\rho/\varepsilon - 2\nu}}{\rho^{1/2}\left(e^{i2\theta} + 1 + \frac{4\varepsilon\nu}{\rho} + \frac{4\varepsilon^{2}\nu^{2}}{\rho^{2}}\right)} \\
		&\sim \dfrac{\varepsilon^{1/2-\beta}e^{2\varepsilon\nu\left(\frac{\rho}{\varepsilon}+2\nu+\beta-1/2\right)/\rho} e^{-i\theta(2N+\beta-2)}e^{-\rho/\varepsilon - 2\nu}}{\rho^{1/2}(e^{i2\theta} + 1)} \\
		&\sim \dfrac{\varepsilon^{1/2-\beta}e^{-i\theta(2N-2+\beta)}e^{-\rho/\varepsilon}}{\rho^{1/2}(e^{i2\theta} + 1)}.
	\end{align*}
	Therefore, the right hand side of \eqref{eqn:remainder_eqn} is
	\begin{align*}
		\sim (-1)^{N}2\sqrt{2\pi}\dfrac{\varepsilon^{1/2-\beta}e^{-i\theta(2N-2+\beta)}e^{-\rho/\varepsilon}}{\rho^{1/2}(e^{i2\theta} + 1)} \left[\cosh(\kappa e^{-i\theta}) - 1\right]f_{0},
	\end{align*}
	as $\varepsilon \to 0$.
	\par
	With this, we consider the remainder equation as
	\begin{equation} \label{eqn:remainder_eqn_2}
		\begin{aligned}
			&\Delta R_{N} - \varepsilon^{2}(-1 + 3\phi_{0}^{2})R_{N} \\
			\sim &(-1)^{N}2\sqrt{2\pi}\dfrac{\varepsilon^{1/2-\beta}e^{-i\theta(2N-2+\beta)}e^{-\rho/\varepsilon}}{\rho^{1/2}(e^{i2\theta} + 1)} \left[\cosh(\kappa e^{-i\theta}) - 1\right]f_{0}.
		\end{aligned}
	\end{equation}
	If we now use the ansatz $R_{N} = e^{\pm i \kappa (x-\zeta)/\varepsilon}S_{N}(x)$, then the left hand side of \eqref{eqn:remainder_eqn_2} becomes
	\begin{align*}
		e^{\pm i\kappa(x-\zeta)/\varepsilon}\Bigg\{&\left[2(\cos\kappa-1)S_{N} + \varepsilon^{2} \cos\kappa S_{N}'' + O(\varepsilon^{4})\right]\\
		-& \varepsilon^{2}(-1+3\phi_{0}^{2})S_{N}\Bigg\}.
	\end{align*}
	At leading order, we get
	\begin{align*}
		\varepsilon^{2}\left[S_{N}''-(-1+3\phi_{0}^{2})S_{N}\right]e^{\pm i\kappa(x-\zeta)/\varepsilon}.
	\end{align*}
	Thus, we would get the equation for $S_{N}(x)$ at leading order to be
	\begin{equation} \label{eqn:SN_eqn}
		\begin{aligned}
			S_{N}''\  &-\  (-1 + 3\phi_{0}^{2})S_{N} \\
			\sim \ &2\sqrt{2\pi}\varepsilon^{-3/2-\beta}(-1)^{N}{e^{-i\theta(2N-2+\beta)}\rho^{-1/2}} \dfrac{\cosh(\kappa e^{-i\theta}) - 1}{(e^{i2\theta} + 1)} \\
			&f_{0} \exp[\mp i \rho e^{i\theta}/\varepsilon - \rho/\varepsilon].
		\end{aligned}
	\end{equation}
	Notice that the right-hand side is no longer exponentially small in $\varepsilon$ if $\theta = \pm \pi/2$; it would be algebraic in $\varepsilon$. This is the Stokes line where the Stokes phenomenon would occur. At the poles $\zeta$ with $\text{Im}\zeta > 0$, the Stokes line is given by $\theta = -\pi/2$ and accordingly it is given by $\theta = \pi/2$ at $\zeta$ where $\text{Im}\zeta < 0$. 
	\par
	By symmetry, the contribution from $\bar{\zeta}$ is the complex conjugate of the contribution from $\zeta$. Thus, it suffices to analyze the case when $\text{Im}\zeta > 0$, in which case $R_{N} = e^{-i\kappa (x-\zeta)/\varepsilon}S_{N}(x)$.
	\par
	To analyze the behavior near the Stokes line, we define the inner angle variable $\tilde{\theta}$ as 
	\begin{equation} \label{eqn:inner_theta}
		\tilde{\theta} = \dfrac{\theta - \left(-\frac{\pi}{2}\right)}{\sqrt{\varepsilon}}.
	\end{equation}
	With the variable, we may approximate the right-hand side of \eqref{eqn:SN_eqn} (see the appendix) and obtain the equation for $S_{N}$ in leading order to be
	\begin{equation} \label{eqn:SN_eqn_inner_theta}
		\begin{aligned}
			S_{N}'' &- (-1 + 3\phi_{0}^{2})S_{N} \\
			\sim &-i\sqrt{\frac{\pi}{2}}\rho^{-1/2}e^{-i\beta\pi/2}\varepsilon^{-1-\beta}\kappa^{2}\tilde{\theta}e^{-\rho\tilde{\theta}^{2}/2}f_{0}.
		\end{aligned}
	\end{equation}
	In terms of $\tilde{\theta}$, the derivative becomes
	\begin{align*}
		\partial_{x} = \frac{\kappa}{\rho}e^{-i\sqrt{\varepsilon}\tilde{\theta}}\varepsilon^{-1/2}\partial_{\tilde{\theta}}.
	\end{align*}
	Thus, $S_{N}'' \sim \frac{\kappa^{2}}{\rho^{2}}\varepsilon^{-1}S_{N,\tilde{\theta}\tilde{\theta}}$. Since the derivative term is more dominant, thus we have a leading order
	\begin{align} \label{eqn:SN_eqn_2}
		S_{N,\tilde{\theta}\tilde{\theta}} &\sim -i\sqrt{\frac{\pi}{2}}\rho^{3/2} e^{i\beta\pi/2} \varepsilon^{-\beta}\tilde{\theta}e^{-\rho\tilde{\theta}^{2}/2} f_{0}.
	\end{align}
	If we now write $S_{N} = \varepsilon^{-\beta}f_{0}(x) \hat{S}_{N}(\tilde{\theta})$, since $f_{0}$ is slowly varying with $\tilde{\theta}$, we have that
	\begin{align*}
		S_{N,\tilde{\theta}\tilde{\theta}} \sim \varepsilon^{-\beta} f_{0} \tilde{S}_{N,\tilde{\theta}\tilde{\theta}}.
	\end{align*}
	Therefore, we have eliminated $f_{0}$ and we are left with the leading order equation for $\hat{S}_{N}$ as
	\begin{align} \label{eqn:SN_hat_eqn}
		\hat{S}_{N,\tilde{\theta}\tilde{\theta}} &\sim -i\sqrt{\dfrac{\pi}{2}} e^{i\beta\pi/2}\rho^{3/2}\tilde{\theta}e^{-\rho\tilde{\theta}^{2}/2}.
	\end{align}
	Integrating twice and imposing that $\hat{S}_{N} \to 0$ as $\tilde{\theta} \to -\infty$, we have
	\begin{align} \label{eqn:SN_hat}
		\hat{S}_{N}(\tilde{\theta}) \sim \dfrac{i\pi}{2}e^{i\beta\pi/2} \text{erfc}\left(-\tilde{\theta}\sqrt{\rho/2}\right),
	\end{align}
	where $\text{erfc}(x)$ is the complementary error function
	\begin{align*}
		\text{erfc}(x) &= \dfrac{2}{\pi}\int_{x}^{\infty}e^{-y^{2}}dy.
	\end{align*}
	It is now readily seen that as the Stokes line is crossed from $\text{Re}(x) < 0$ to $\text{Re}(x) > 0$, we have a jump of size
	\begin{align*}
		\hat{S}_{N}(\theta \to (-\pi/2)^{+}) - \hat{S}_{N}(\theta \to (-\pi/2)^{-}) = i\pi e^{i\beta\pi/2},
	\end{align*}
	owing to the pole $\zeta$. Note that this jump happens smoothly as the Stokes line is crossed. However, the change of value in $\hat{S}_{N}$ happens in the scale of $\tilde{\theta}$, i.e, when $\text{arg}(x-\zeta) - (-\pi/2) = O(\varepsilon^{1/2})$, causing it to appear rapidly.
	\par
	With this, the remainder term due to the pole $\zeta$ with $\text{Im}(\zeta) > 0$ is
	
	\begin{equation} \label{eqn:R_N}
		\begin{aligned}
			R_{N} &\sim e^{-i\kappa(x-\zeta)/\varepsilon} S_{N}(x)\\
			&\sim \frac{i\pi}{2}e^{i2\pi\left(n_{0}+\frac{\zeta}{\varepsilon}\right)}\varepsilon^{-4}\text{erfc}\left(-\tilde{\theta}\sqrt{\frac{\rho}{2}}\right)\Lambda_{1}G(x).
		\end{aligned}
	\end{equation}
	Recall that $\kappa = 2\pi$, $\beta = 4$, $\rho = |\kappa(x-\zeta)|$, $f_{0}(x) = \Lambda_{1} G(x)$ and $\tilde{\theta}$ as defined in \eqref{eqn:inner_theta}.
	\par
	From \eqref{eqn:R_N}, we see that the dominant contribution to the remainder comes from the pole $\zeta$ nearest to the real axis. In this case, they are the poles $\zeta_{1} = i\pi/\sqrt{2}$ and its conjugate $\overline{\zeta}_{1} = -i\pi/\sqrt{2}$ which are equidistant from the real axis.
	\par
	Therefore, from the poles $\zeta$ with $\text{Im}\zeta > 0$, crossing the Stokes line amounts to the remainder $R_{N}$ experiencing a jump of size
	\begin{align*}
		\sim i \pi \varepsilon^{-4} e^{i2\pi\left(n_{0} + \frac{\zeta}{\varepsilon}\right)} \Lambda_{1} G(x),
	\end{align*}
	which is of exponentially small order in $\varepsilon$.
	\par
	As noted earlier, the dominant contribution in the remainder comes from the poles closest to the real axis, i.e., $\zeta = \zeta_{1} = i\pi/\sqrt{2}$ and its conjugate. Combining the leading order term $\phi_{0}(x)$ with the dominant term of $R_{N}$ from both $\zeta_{1}$ and $\overline{\zeta}_{1}$, we have
	\begin{align} \label{eqn:phi_asymp_cubic_defocusing}
		\phi(x) \sim \phi_{0}(x) +  K \varepsilon^{-4}{e^{-\sqrt{2}\pi^{2}/\varepsilon}}\text{erfc}\left(-\tilde{\theta}\sqrt{\frac{\rho}{2}}\right)G(x) + \text{c.c.},
	\end{align}
	where $K = \dfrac{i\pi}{2}e^{i2\pi n_{0}}\Lambda_{1}$. Effectively, as $x > 0$, $\phi(x)$ is asymptotically
        \begin{equation} \label{eqn:phi_asymp_x_pos}
            \begin{split}
                \phi(x) \sim\ &\phi_{0}(x) \\
                &+C_{R}\left[\text{Re}\ G(x)\sin(2\pi n_{0})+ \text{Im}\ G(x) \cos(2\pi n_{0})\right]
            \end{split}
        \end{equation}
        where $C_{R} = 2\pi|\Lambda_{1}|\varepsilon^{-4}e^{-\sqrt{2}\pi^{2}/\varepsilon}$.
	\par
	For \eqref{eqn:phi_asymp_cubic_defocusing} to be a solution of the stationary DNLS \eqref{eqn:dnls_main_cubic_defocusing_stationary}, the remainder needs to vanish as $x \to \infty$. However, from \eqref{eqn:f0_G}, $G(x)$ grows exponentially as $x \to \infty$, in fact, as $x \to \infty$, the behaviour of $G(x)$ is
	\begin{align*}
		G(x) \sim \dfrac{1}{8}e^{\sqrt{2}x}.
	\end{align*}
	Thus, the exponentially small tail of $\phi(x)$ in \eqref{eqn:phi_asymp_x_pos} will eventually grow as
    \begin{align*}
        \sim \frac{1}{8}e^{\sqrt{2}x}\sin(2\pi{n_{0}})2\pi|\Lambda_{1}|\varepsilon^{-4}e^{-\sqrt{2}\pi^{2}/\varepsilon} 
    \end{align*}
	This growing term can be eliminated if we choose $n_{0}$ 
    such that
	\begin{align}
		\sin(2\pi n_{0}) = 0.
	\end{align}
	Thus, for $\phi(x)$ to be bounded, it must be that $n_{0} = 0$ or $n_{0} = 1/2$ modulo 1. These two values of $n_{0}$ correspond respectively to the onsite solution $(n_{0} = 0)$ and intersite solution $(n_{0} = 1/2)$.
	
	\subsection{Linear Stability}

Next, we study the stability of the dark solitons obtained above, i.e., \eqref{eqn:phi_asymp_cubic_defocusing} with $n_0$ being one of the two values $n_0 = 0$ or $n_0 = 1/2$. Consider the eigenvalue problem \eqref{eqn:dnls_defocusing_evp}. The eigenvalue equation can be considered as a single equation 
	\begin{align} 
		L_{-}L_{+}u = -\lambda^{2} u.
	\end{align}
	
	Assuming that $\lambda$ is small, we may expand $u$ as
	\begin{align}
		u &= u_{0} + \lambda^{2}u_{1} + \lambda^{4}u_{2} + \dots. \label{eqn:eigvec_expansion_cubic}
	\end{align}
	At $O(1)$, we have
	\begin{align*}
		L_{-}L_{+}u_{0} = 0.
	\end{align*}
	This can be solved by observing that $L_{-}\tilde{\phi} = 0$ for any $n_{0}$. 
	Now, we let $x_{0} = \eps n_{0}$ be the scaled center of the solution. 
	Differentiating $L_{-}\tilde{\phi} = 0$ with respect to $x_{0}$, we obtain $L_{+}(\partial_{x_{0}}\tilde{\phi}) = 0$. Therefore, we get that $L_{-}L_{+}(\partial_{x_{0}}\tilde{\phi}) = 0$. Thus, we set $u_{0} = \partial_{x_{0}}\tilde{\phi}$, with $n_{0}$ taking either the value $0$ or $1/2$.
	\par
	Notice that as $x \to \infty$, the leading order behavior of $u_{0}$ is
	\begin{align}
		u_{0} &\sim \dfrac{\pi^{2}e^{-\sqrt{2}\pi^{2}/\varepsilon}|\Lambda_{1}|}{2\varepsilon^{4}}\cos(2\pi n_{0})e^{\sqrt{2}x},
	\end{align}
	which has an exponentially small growing term owing to the $e^{\sqrt{2}x}$ term. We will see that this growing term will be eliminated at the next order by $u_{1}$.
	\par
	At order $O(\lambda^{2})$, we have the equation
	\begin{align} \label{eqn:LmLp_eqn_O(lamb2)_defocusing}
		L_{-}L_{+}u_{1} &= -u_{0},
	\end{align}
	which can be solved successively by solving the coupled equations
	\begin{align}
		L_{-}w_{1} &= -u_{0}, \label{eqn:Lm_eqn_O(lamb2)_defocusing} \\
		L_{+}u_{1} &= w_{1}. \label{eqn:Lp_eqn_O(lamb2)_defocusing}
	\end{align}
	The first of the two equations \eqref{eqn:Lm_eqn_O(lamb2)_defocusing} can be solved by writing it as
	\begin{align} \label{eqn:w1_eqn_defocusing}
		\Delta w_{1} + \varepsilon^{2}(w_{1} - \tilde{\phi}^{2}w_{1}) = \varepsilon^{2}u_{0}.
	\end{align}
	Now, if we suppose the ansatz $w_{1} = e^{ikn}h(x)$, upon substitution to \eqref{eqn:w1_eqn_defocusing}, we would get up to $\mathcal{O}(\varepsilon^{3}h'''(x))$
	\begin{align*}
		\varepsilon^{2}u_{0} 
		= \ &2(\cos k - 1)h(x) + \varepsilon 2i \sin(k) h'(x) \\
		&+ \varepsilon^{2}\left[\cos(k)h''(x) + (1 - \tilde{\phi}^{2})h(x))\right].
	\end{align*}
	For $h(x)$ to be nonzero, it follows that $\cos k - 1 = 0$, that is, $k$ is a multiple of $2\pi$. Thus, the order $O(\varepsilon)$ equation is automatically satisfied. At order $O(\varepsilon^{2})$, we obtain the equation for $h(x)$ to be
	\begin{align} \label{eqn:h_defocusing}
		h''(x) + (1-\tilde{\phi}^{2})h(x) &= u_{0}.
	\end{align}
	Having chosen $n_{0}$ to be one of the two values that eliminate the growing term in $\phi$ in the far field, the remainder term in $\tilde{\phi}$ remains exponentially small and thus is neglected from the calculations. In effect, we are taking $\tilde{\phi}$ to be $\phi_{0}(x)$ with $n_{0} = 0$ or $n_{0} = 1/2$.
	\par
	Similarly, since the growing term in $u_{0}(x)$ is exponentially small, we also neglect it in the calculation of $h(x)$, thus $u_{0}$ can be regarded as $u_{0} = - \phi_{0}'(x)$. We thus get a solution for $h(x)$ as
	\begin{align*}
		h(x) = -\dfrac{1}{2} x\phi_{0}(x).
	\end{align*}
	Next, we solve the second equation \eqref{eqn:Lp_eqn_O(lamb2)_defocusing}, which now becomes
	\begin{align*}
		L_{+} u_{1} &= -\dfrac{1}{2} x \phi_{0}(x).
	\end{align*}
	Again, using the ansatz $u_{1} = e^{ikn}f(x)$, we get that nontrivial $f(x)$ implies that $k$ is a multiple of $2\pi$. As such, we will get an equation for $f(x)$ at $O(\varepsilon^{2})$ to be
	\begin{align} \label{eqn:f_eqn_defocusing}
		f''(x) + (1-3\phi_{0}^{2})f(x) &= \dfrac{1}{2} x\phi_{0}(x).
	\end{align}
	The homogeneous version of this equation has a solution $\phi_{0}'(x)$. By reduction of order, we may get the solution to \eqref{eqn:f_eqn_defocusing} of the form $f(x) = \phi_{0}'(x)v(x)$, where $v(x)$ satisfies
	\begin{align*}
		v'(x) = &\dfrac{1}{[\phi_{0}'(x)]^{2}} \dfrac{1}{2}\int_{-\infty}^{x} \xi\phi_{0}(\xi)\phi_{0}'(\xi) d\xi \\
		= &\dfrac{1}{{2\sqrt{2}}[\phi_{0}'(x)]^{2}} \left[\tanh\left(\frac{x}{\sqrt{2}}\right) - \frac{x}{\sqrt{2}}\sech^{2}\left(\frac{x}{\sqrt{2}}\right) + 1\right].
	\end{align*}
	As $x \to \infty$, we have
	\begin{align*}
		v'(x) 
		&\sim \dfrac{1}{8\sqrt{2}}e^{2\sqrt{2}x}.
	\end{align*}
	Thus, the far field behavior of $u_{1}$ is
	\begin{align*}
		u_{1} = f(x) &\sim \dfrac{1}{8\sqrt{2}}e^{\sqrt{2}x}.
	\end{align*}
	Substituting this into the eigenvector expansion, we have as $x \to \infty$
	\begin{align*}
		u &= u_{0} + \lambda^{2}u_{1} + \dots \\
		\sim &-\varepsilon\phi_{0}'(x) + \left[\dfrac{\pi^{2}e^{-\sqrt{2}\pi^{2}/\varepsilon}|\Lambda_{1}|}{2\varepsilon^{4}}\cos(2\pi n_{0}) + \lambda^{2}\frac{1}{8\sqrt{2}}\right]e^{\sqrt{2}x} \\
		&+ \dots
	\end{align*}
	Eliminating the growing term gives an equation for $\lambda$ as 
	\begin{align} 
		\label{eqn:dnls_dark_ev_approximation}
		\lambda^{2} &\sim -{4\sqrt{2}\pi^{2}|\Lambda_{1}|\varepsilon^{-5}e^{-\sqrt{2}\pi^{2}/\varepsilon}} \cos(2\pi n_{0}). 
	\end{align}
	
	The asymptotic result Eq. \eqref{eqn:dnls_dark_ev_approximation} implies that the intersite solution $(n_{0} = 1/2)$ possesses a pair of (exponentially small) real eigenvalues and is thus linearly unstable. According to the expression, the onsite solution $(n_{0}=0)$ would possess a pair of imaginary eigenvalues and hence is stable. However, this contrasts with the fact that onsite solitons are unstable due to two pairs of complex eigenvalues \cite{Johansson2004}. 
	\par
	Although direct linear stability analysis fails to predict the oscillatory instability of the onsite dark soliton, here we offer an alternative approach to obtain the qualitative behavior of the instability based upon an eigenvalue counting argument as employed by Pelinovsky and Kevrekidis \cite{pelinovsky2008stability, Pelinovsky2007}. The argument provides a way of counting the eigenvalues of $\mathcal{L}$ by considering the number of negative eigenvalues of $L_{-}$ and $L_{+}$. These eigenvalue counts are related to each other as given by the following lemma.
	
	\begin{lemma} \label{lemma:eigval_count}
		Suppose that $L_{\pm}$ have trivial kernels in $l^{2}(\Z)$ and that $L_{\pm}$ each possesses $n_{\pm}$ negative eigenvalues. Furthermore, assume that all 
        eigenvalues of the spectral problem \eqref{eqn:dnls_defocusing_evp} are algebraically simple. Then, $\mathcal{L}$ has $N_{c}$ complex eigenvalues in the first quadrant, $N_{i}^{-}$ purely imaginary eigenvalues with positive imaginary part satisfying $\langle v, L_{+}^{-1}v \rangle \leq 0$, $N_{r}^{+}$ real positive eigenvalues for which $\langle v, L_{+}^{-1}v \rangle \geq 0$, and $N_{r}^{-}$ real positive eigenvalues satisfying $\langle v, L_{+}^{-1}v \rangle \leq 0$. These eigenvalue counts are related through the equations:
		\begin{align}
			N_{r}^{-} + N_{i}^{-} + N_{c} &= n_{+}, \label{eqn:count_eigval_1}\\
			N_{r}^{+} + N_{i}^{-} + N_{c} &= n_{-}. \label{eqn:count_eigval_2}
		\end{align}
	\end{lemma}
	\begin{proof}
		Since $|\tilde{\phi}|^{2} \to 1$ exponentially fast as $|x| \to \infty$, by Weyl's Essential Spectrum Lemma, the essential spectra of $L_{+}$ and $L_{-}$ are respectively bounded below by $2$ and $0$.
		Now, since by assumption, the kernel of $L_{+}$ is empty in $l^{2}(\Z)$, we may consider the eigenvalue equation \eqref{eqn:dnls_defocusing_evp} as the generalized eigenvalue equation
		\begin{equation} \label{eqn:eigval_eqn_2}
			L_{-}u = \gamma L_{+}^{-1} v, \quad \gamma = -\lambda^{2}.
		\end{equation}
		Furthermore, we can shift this equation, rewriting it as
		\begin{equation} \label{eqn:eigval_eqn_3}
			\left(L_{-} + \delta L_{+}^{-1}\right)v = (\gamma + \delta) L_{+}^{-1} v,
		\end{equation}
		for sufficiently small $\delta > 0$.
		
		Now, properties P1 and P2 of \cite{chugunova2010count} are satisfied by $L_{\pm}$, with $L_{\pm}$ in this case taking the roles of $L_{\mp}$ in \cite{chugunova2010count}. In the context of \cite{chugunova2010count}, the constrained space of interest in our case is the whole space $l^{2}(\Z)$ since $\ker L_{+}$ is empty by assumption. Thus, we have that Theorem 3 of \cite{chugunova2010count} applies. As a result, we obtain the relations
		\begin{align}
			N_{r}^{-} + N_{i}^{-} + N_{c} &= \dim\left(\mathcal{H}^{-}_{L_{+}^{-1}}\right), \\
			N_{r}^{+} + N_{i}^{-} + N_{c} &= \dim\left(\mathcal{H}^{-}_{L_{-}+\delta L_{+}^{-1}}\right),
		\end{align}
		where $\mathcal{H}^{-}_{A}$ denotes the negative invariant subspaces of $l^{2}(\Z)$ associated with the operator $A$. By the assumptions satisfied by $L_{\pm}$, particularly the fact that the spectrum of $L_{+}$ being bounded away from zero and having a discrete spectrum of eigenvalues with trivial kernel, it follows that the negative invariant subspace of $L_{+}^{-1}$ corresponds exactly to the negative invariant subspace of $L_{+}$, which dimension is given by the number of negative eigenvalues of $L_{+}$, thus we have $\dim(\mathcal{H}^{-}_{L_{+}^{-1}}) = n_{+}$.
		
		Finally, we need to show that \[\dim(\mathcal{H}^{-}_{L_{-}+\delta L_{+}^{-1}}) = \dim(\mathcal{H}^{-}_{L_{-}}) = n_{-}.\] 
		This can be demonstrated as follows.
		The inequality $\dim(\mathcal{H}^{-}_{L_{-}}) \leq \dim(\mathcal{H}^{-}_{L_{-} + \delta L_{+}^{-1}})$ follows from the continuity of the eigenvalues and the relative compactness of $L_{+}^{-1}$ with respect to $L_{-}$. The key issue is whether, for small $\delta > 0$, an edge bifurcation causes an eigenvalue of $L_{-} + \delta L_{+}^{-1}$ to emerge from the lower edge of the essential spectrum of $L_{-}$. To investigate this, consider the eigenvalue problem
		\begin{equation}\label{eqn:eigval_eqn_Lm+dLp-1}
			(L_{-} + \delta L_{+}^{-1})w = \omega w.
		\end{equation}
		This equation can also be written as
		\begin{equation}
			(L + \delta B)w = \omega w,
		\end{equation}
		where
		\begin{align}
			L &= L_{-} + \delta L_{+,\infty}^{-1}, \\ 
			B &= L_{+}^{-1}\left(L_{+,\infty} - L_{+}\right)L_{+,\infty}^{-1} \\
			&= L_{+}^{-1}\left(3 - 3|\tilde{\phi}|^{2}\right)L_{+,\infty}^{-1}.
		\end{align}
		Here, $L_{+,\infty} = -\dfrac{1}{\eps^{2}}\Delta + 2$ represents the limiting form of the operator $L_{+}$ as $|x| \to \infty$. The operator $B$ constitutes a relatively compact perturbation to $L$, and the essential spectrum of $L$ is bounded below by $\delta/2 > 0$. 
		
		By applying the theory of edge bifurcations \cite{kapitula_stanstede2004_eigenvalues_evans}, any edge bifurcation that occurs when $\delta \neq 0$ and originates at the lower edge of the essential spectrum of $L$ would appear as
		\begin{equation}
			\omega(\delta) = \frac{\delta}{2} - a \delta^{2} + \mathcal{O}(\delta^{3}),
		\end{equation}
		where $a > 0$ is a constant. For sufficiently small $\delta > 0$, $\omega(\delta) > 0$, which implies that the number of negative eigenvalues of $L_{-} + \delta L_{+}^{-1}$ remains unchanged. Consequently, $\dim(\mathcal{H}^{-}_{L_{-}+\delta L_{+}^{-1}}) = \dim(\mathcal{H}^{-}_{L_{-}}) = n_{-}$ for small $\delta > 0$.
		
		Thus, we have established that \eqref{eqn:count_eigval_1} and \eqref{eqn:count_eigval_2} hold.
	\end{proof}
    
In the context of the dark solitons, we have the following theorem:
\begin{theorem}\label{thm:eigval_count}
    In the limit $\epsilon\to0$, the onsite and off-site dark solitons have $n_+=1$ and $0$, respectively. On the other hand, both have $n_-=1$. This implies that the off-site soliton must have one real eigenvalue. Assuming that there is no eigenvalue embedded in the continuous spectrum of the operator $\mathcal{L}$, i.e., $N_i^-=0$, the onsite soliton must have one complex eigenvalue.
\end{theorem}

The operator $L_{-}$ has an empty kernel in $l^{2}(\Z)$, since $L_{-}\tilde{\phi} = 0$ and that $\tilde{\phi} \in l^{\infty}(\Z)$ but $\tilde{\phi} \notin l^{2}(Z)$. Moreover, $\tilde{\phi}$ has one sign change; therefore, Sturm-Liouville theory gives that $L_{-}$ possesses one negative eigenvalue, thus $n_{-} = 1$. We have also established that $L_{+}$ has an empty kernel in $l^{2}(\Z)$ since although $L_{+}u_{0} = L_{+}\tilde{\phi}_{n_{0}} = 0$, $u_{0}$ has a growing tail as $x \to \infty$, and thus $u_{0} \notin l^{2}(\Z)$. It remains now to compute the number of negative eigenvalues of $L_{+}$. This may be done by a similar analysis as the direct linear stability as above. In the continuum limit, $L_{+}$ has a zero eigenvalue due to the translation invariance of the system in this limit. However, in the discrete case, this translation invariance is broken, and we may consider the bifurcation of the zero eigenvalue as before. We consider the eigenvalue equation
	\begin{equation} \label{eqn:Lp_ev_eqn_dark}
		L_{+}u = \alpha u,
	\end{equation}
	where now we expand $u$ as
	\begin{equation} \label{eqn:u_expansion_Lp}
		u = u_{0} + \alpha u_{1} + \dots \quad (\alpha \ll 1),
	\end{equation}
	where we take $u_{0} = \partial_{x_{0}}\tilde{\phi}$ as before. This ensures that upon substitution of \eqref{eqn:u_expansion_Lp} to \eqref{eqn:Lp_ev_eqn_dark}, the equation at $O(1)$ is automatically satisfied. At order $O(\alpha)$, we have the equation
	\begin{equation} \label{eqn:Lp_ev_eqn_O(lambda)}
		L_{+} u_{1} = u_{0}.
	\end{equation}
	As before, assuming that $u_{1} = u_{1}(x)$, this leads to the equation
	\begin{equation} \label{eqn:Lp_u1_eqn}
		u_{1}''(x) - \left(1 - 3\phi_{0}^{2}\right)u_{1}(x) = \phi_{0}'(x).
	\end{equation}
	Solving this for $u_{1}(x)$, it is found that as $x \to \infty$, $u_{1}(x)$ has a growing tail as
	\begin{equation} \label{eqn:Lp_u1_tail}
		u_{1}(x) \sim \dfrac{1}{6\sqrt{2}}e^{\sqrt{2}x}.
	\end{equation}
	Therefore, as $x \to \infty$, we have that
	\begin{equation}
		\begin{split}
			u &= u_{0} + \alpha u_{1} \\
			&\sim \left[\dfrac{\pi^{2} e^{-\sqrt{2}\pi^{2}/\varepsilon}|\Lambda_{1}|}{2\eps^{5}} \cos(2\pi n_{0}) + \alpha\dfrac{1}{6\sqrt{2}}\right] e^{\sqrt{2}x}.
		\end{split}
	\end{equation}
	To counterbalance the growing tail of $u_{0}$, we have that as $\varepsilon \to 0$, $\alpha$ is
	\begin{equation} \label{eqn:Lp_ev_asymp}
		\alpha \sim -\dfrac{3\sqrt{2}\pi^{2}e^{-\sqrt{2}\pi^{2}/\varepsilon}|\Lambda_{1}|}{\varepsilon^{5}} \cos(2\pi n_{0}).
	\end{equation}
	We may now deduce that the zero eigenvalue of $L_{+}$ bifurcates to a negative eigenvalue in the case of the onsite dark soliton. This is in contrast to the case of the intersite dark soliton, where it bifurcates to a positive eigenvalue. Moreover, by the positivity of each of the corresponding eigenvectors, $L_{+}$ has no negative eigenvalue in the case of the intersite dark soliton. Thus, we have that $n_{+} = 1$ for the onsite dark soliton and $n_{+} = 0$ for the intersite variant.

    From these results, we may now count the eigenvalues of $\mathcal{L}$ for the respective cases of onsite and intersite dark solitons. For intersite dark solitons, Lemma \ref{lemma:eigval_count} gives 
	\begin{align}
		N_{r}^{-} + N_{i}^{-} + N_{c} &= 0, \\
		N_{r}^{+} + N_{i}^{-} + N_{c} &= 1,
	\end{align}
	from which it is immediately seen that $N_{i}^{-} = N_{c} = 0$ and $N_{r}^{+} = 1$, giving one real eigenvalue and thus confirming our preceding results from the linear stability analysis. On the other hand, in the case of onsite dark solitons, we have
	\begin{align}
		N_{r}^{-} + N_{i}^{-} + N_{c} &= 1, \\
		N_{r}^{+} + N_{i}^{-} + N_{c} &= 1.
	\end{align}
	In the continuum limit, the essential spectrum of $\mathcal{L}$ occupies the whole imaginary axis. Assuming that there are no embedded eigenvalues as $\varepsilon \to 0$, i.e., we rule out any imaginary eigenvalues of $\mathcal{L}$, we have $N_{i}^{-} = 0$. Now, suppose to the contrary that $N_{r}^{-}$ and $N_{r}^{+}$ are the two quantities which are equal to 1. This implies the existence of two real eigenvalues in the first quadrant near the continuum limit. These pairs of real eigenvalues must then converge to the zero eigenvalues in the continuum limit, each converging to the zero eigenvalues owing to phase and translation invariance, respectively. However, the phase-invariant eigenvalue remains zero in the discrete case. Therefore, only one pair of eigenvalues can bifurcate from zero. Thus, we cannot have two pairs of real eigenvalues in the near-continuum limit case. This leaves us with the only possibility that $N_{c} = 1$, i.e., a complex eigenvalue exists in the first quadrant, leading to a quartet of complex eigenvalues, which aligns with the numerical results.

	\begin{figure*}
		\centering
		\includegraphics[width=0.9\linewidth]{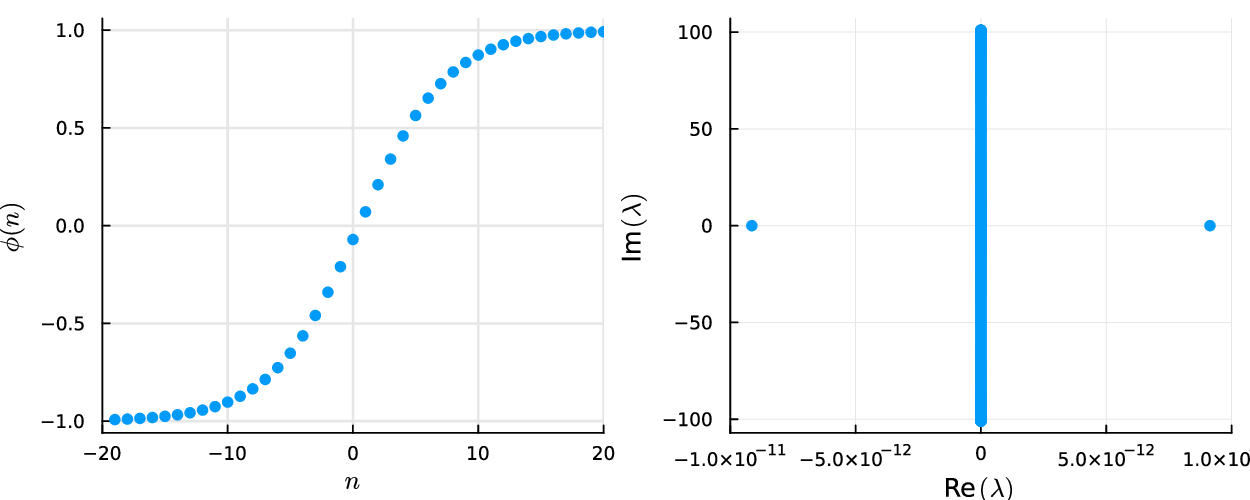}
		\caption{{An intersite dark soliton for $\varepsilon=0.2$. The left and right panels show the stationary solution and corresponding spectra in the complex plane. The inset shows a pair of discrete spectra on the real axis, indicating the instability of the soliton.}}
		\label{fig:dark-sol-intersite}
	\end{figure*}
	
	\begin{figure*}
		\centering
		\includegraphics[width=0.8\linewidth]{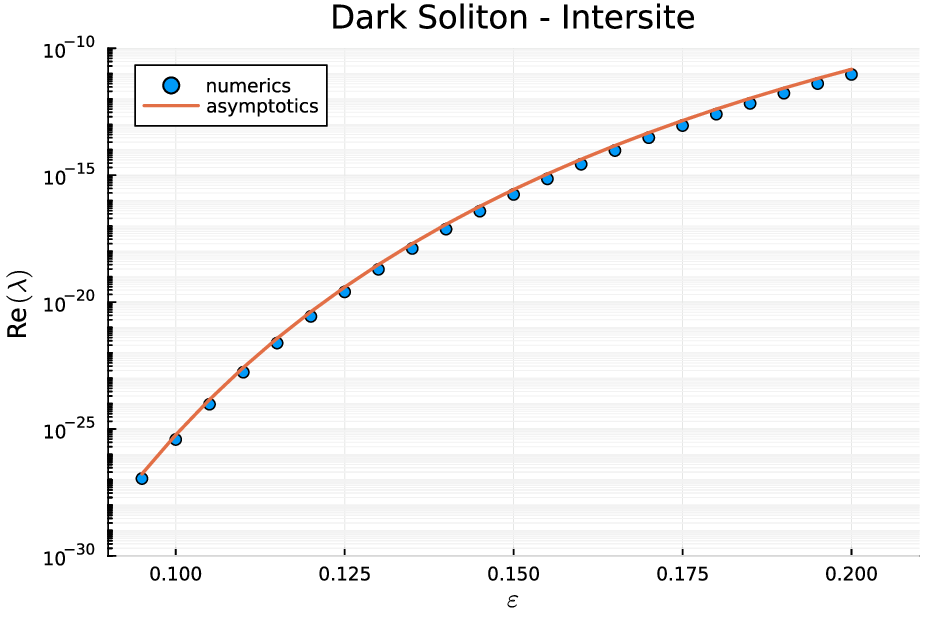}
		\caption{{The numerically computed critical eigenvalue of the intersite dark soliton for varying $\varepsilon$ and our exponential asymptotics Eq.\ \eqref{eqn:dnls_dark_ev_approximation}.
		}}
		\label{fig:dark-eigvals-intersite}
	\end{figure*}
	
	\begin{figure*}
		\centering
		\includegraphics[width=0.9\linewidth]{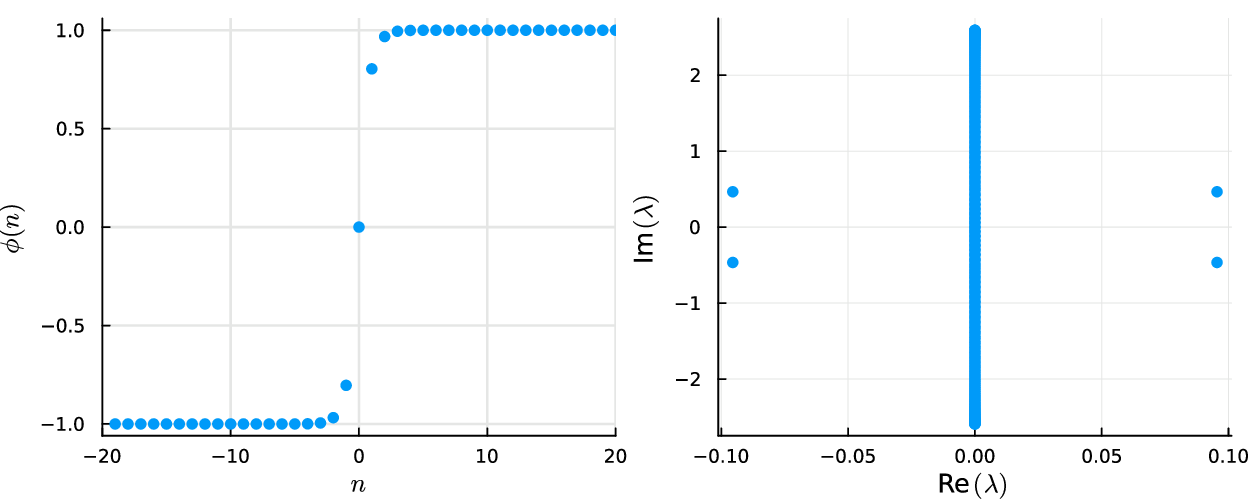}
		\caption{
			{Similar to Fig.\ \ref{fig:dark-sol-intersite}, but for an onsite dark soliton with $\varepsilon=1.5$. 
				The right panel displays the corresponding spectrum with pairs of imaginary eigenvalues, demonstrating the oscillatory instability of the soliton.
		} }
		\label{fig:dark-sol-onsite}
	\end{figure*}
	
	\begin{figure*}
		\centering
		\includegraphics[width=0.8\linewidth]{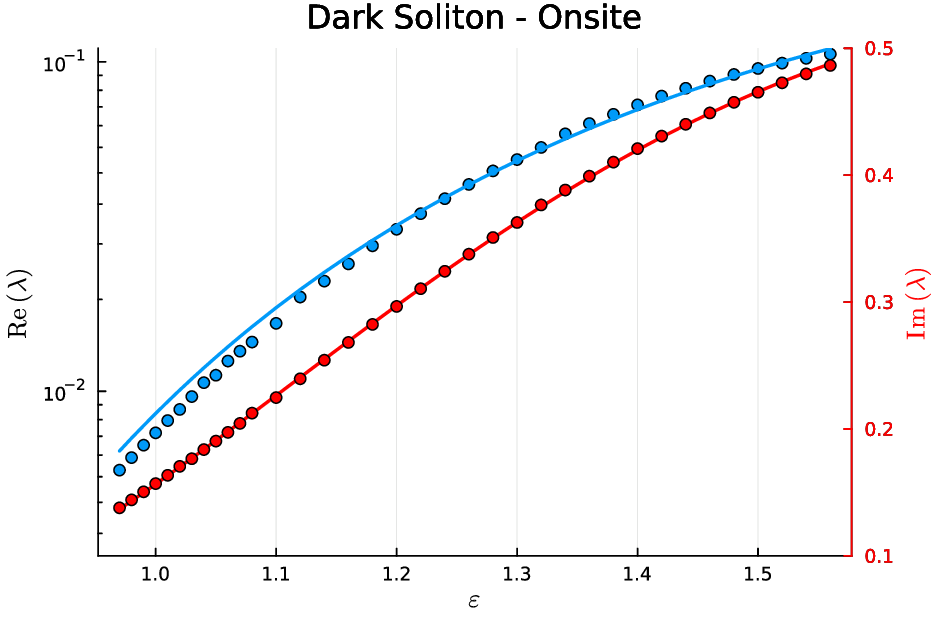}
		\caption{{The critical eigenvalues of the onsite dark soliton for varying $\varepsilon$. We are not able to capture the instability analytically. 
				The solid lines fit based on Eq.\ \eqref{eqn:dark_onsite_fitting}}. }
		\label{fig:dark-eigvals-onsite}
	\end{figure*}
	
	\section{Bright Solitons}
	\label{sec3}
	
We now turn to the case of the focusing nonlinearity
	\begin{equation} \label{eqn:dnls_main_cubic_focusing}
		i\partial_{t}\phi = -\dfrac{C}{2}\Delta \phi +\mu \phi - |\phi|^{2}\phi.
	\end{equation}
Again, 
we consider $\mu = 1$ without loss of generality. We are interested in time-independent bright soliton solutions.

To analyze the stability of the steady-state solution, we perform the same analysis as before. Linearizing around the solution, which we denote as $\tilde{\phi}$, we will get the eigenvalue equation
	\begin{align}\label{eqn:dnls_focusing_evp}
		\lambda \begin{bmatrix}
			u \\ v
		\end{bmatrix} = \mathcal{L}\begin{bmatrix}
			u \\ v
		\end{bmatrix},
	\end{align}
	where now
	\begin{align*}
		\mathcal{L} &=   \begin{bmatrix}
			O & L_{-} \\ -L_{+} & O
		\end{bmatrix}, \\
		L_{-} &= -\dfrac{1}{\varepsilon^{2}}\Delta + 1 - |\tilde{\phi}|^{2}, \quad 
		L_{+} = -\dfrac{1}{\varepsilon^{2}}\Delta + 1 - 3|\tilde{\phi}|^{2}.
	\end{align*}

Our main result in this section is stated in the following theorem:
\begin{theorem}\label{thm:brightsoliton}
The focusing DNLS equation \eqref{eqn:dnls_main_cubic_focusing} in the strong coupling limit $C\to\infty$ admits two types of bright solitons of the form
\[
    \phi(x) \sim \sqrt2\sech\left({x}\right) - C_{R} \textnormal{Im}\ G(x) \cos(2\pi n_{0}),
\]
where $x = \varepsilon(n - n_{0})$, $\varepsilon=\sqrt{2/C}$, $C_{R} = 2\pi|\Lambda_{1}|\varepsilon^{-4}e^{-\pi^{2}/\varepsilon}$ with $\Lambda_{1}$ being given by Eqs.\ \eqref{eqn:recurrence_relation_Aj} and \eqref{eqn:Lambda1_focusing}, $G(x)$ is defined in Eq.\ \eqref{eqn:f0_G}, and $n_{0}$ can take the values $0$ or $1/2$ (mod $1$). The first value, $n_0 = 0$, corresponds to the onsite soliton, while the second value, $n_0 = 1/2$, corresponds to the off-site (or intersite) soliton. The off-site soliton is exponentially unstable due to a pair of real-valued linear eigenvalues, whereas the onsite soliton is stable with a pair of purely imaginary eigenvalues. Their leading asymptotic expression is given in Eq.\ \eqref{eqn:dnls_bright_ev_approximation}.
\end{theorem}

\subsection{Time-independent states}
    
We first consider the time-independent equation
	\begin{align} \label{eqn:dnls_main_cubic_focusing_stationary}
		0 &= \Delta \phi - \varepsilon^{2}(  \phi - |\phi|^{2}\phi).
	\end{align}
	A similar analysis as in the case of dark solitons would give that the leading order solution is \
	\begin{align} \label{eqn:dnls_focusing_leading_order_soln}
		\phi_{0}(x) &= \sqrt{2}\sech(x),
	\end{align}
	which now has poles at
	\begin{align}
		x &= i \dfrac{1}{2 }(2k+1)\pi \quad (k \in \Z).
	\end{align}
	Near these poles, $\phi_{0}(x)$ diverges as
	\begin{align}
		\phi_{0}(x) \sim \dfrac{-i\sqrt{2}}{x-\zeta_{k}}.
	\end{align}
	
	Doing the routine calculations for the late order terms and the remainder gives that for $x > 0$
    \begin{equation} \label{eqn:phi_asymp_cubic_focusing}
            \begin{split}
                \phi(x) \sim\ &\phi_{0}(x) \\
                &-C_{R}\left[\text{Re}\ G(x)\sin(2\pi n_{0})+ \text{Im}\ G(x) \cos(2\pi n_{0})\right]
            \end{split}
        \end{equation}
        where $C_{R} = 2\pi|\Lambda_{1}|\varepsilon^{-4}e^{-\pi^{2}/\varepsilon}$ and $G(x)$ is defined as before in \eqref{eqn:f0_G} with $\phi_{0}(x)$ being the leading order bright soliton solution \eqref{eqn:dnls_focusing_leading_order_soln}. Here, the constant $\Lambda_{1}$ is computed similarly as before by determining the appropriate recurrence relation derived from the inner equation near the pole $\zeta$, which in this case is
        \begin{equation} \label{eqn:recurrence_relation_Aj}
		0 = 2\sum_{p=1}^{j+1}\binom{2j+2}{2p}A_{j-p+1} + \sum_{k=0}^{j}\sum_{l=0}^{k}A_{l}A_{k-l}A_{j-k},
	\end{equation}
	with $A_{0} = -i\sqrt{2}$ and the relation
	\begin{align} \label{eqn:Lambda1_focusing}
		\Lambda_{1} = \dfrac{10}{i\sqrt{2}}\lim_{j \to \infty} \dfrac{(-1)^{j+1}(2\pi)^{2j+4}}{\Gamma(2j+4)}A_{j}.
	\end{align}
	Solving the recurrence relation gives $\Lambda_{1}     \approx 2533$.
	
	As before, $G(x)$ also grows exponentially as $x \to \infty$, in this case
	\begin{align*}
		G(x) \sim -\dfrac{1}{4\sqrt{2}}e^{x}.
	\end{align*}
	Thus, the exponentially small remainder term would also grow exponentially as
	\begin{align*}
		\sim \dfrac{1}{4\sqrt{2}}e^{x}\sin(2\pi n_{0})2\pi|\Lambda_{1}| \varepsilon^{-4}e^{-\frac{\pi^{2}}{\varepsilon}},
	\end{align*}
	when $x \to \infty$. Elimination of this exponentially small growing tail gives the same condition as before, that is, $n_{0}$ must also satisfy
	\begin{align}
		\sin(2\pi n_{0}) = 0.
	\end{align}
    Therefore, the exponentially growing remainder term is eliminated if we choose $n_{0} = 0$ or $n_{0} = 1/2$ as before, giving the onsite and off-site solutions, respectively.
	
	\subsection{Linear Stability}
The eigenvalue equation \eqref{eqn:dnls_focusing_evp} can be considered as a single equation as
	\begin{align}
		L_{-}L_{+}u = -\lambda^{2} u.
	\end{align}
	By writing $u$ as an expansion in powers of $\lambda$ and solving in consecutive powers, we find an equation for $\lambda^{2}$ to be
	\begin{align}
		\label{eqn:dnls_bright_ev_approximation}
		\lambda^{2} &\sim -{4\pi^{2}|\Lambda_{1}|}{}\varepsilon^{-5}e^{-\frac{\pi^{2}}{\varepsilon}} \cos(2\pi n_{0}), \quad (\varepsilon \to 0).
	\end{align}
	
	This gives the result that the intersite solution $(n_{0} = 1/2)$ possesses a pair of (exponentially small) real eigenvalues and is thus linearly unstable. In contrast, the onsite solution $(n_{0}=0)$ possesses a pair of imaginary eigenvalues.
	
	\begin{figure*}
		\centering
		\includegraphics[width=0.9\linewidth]{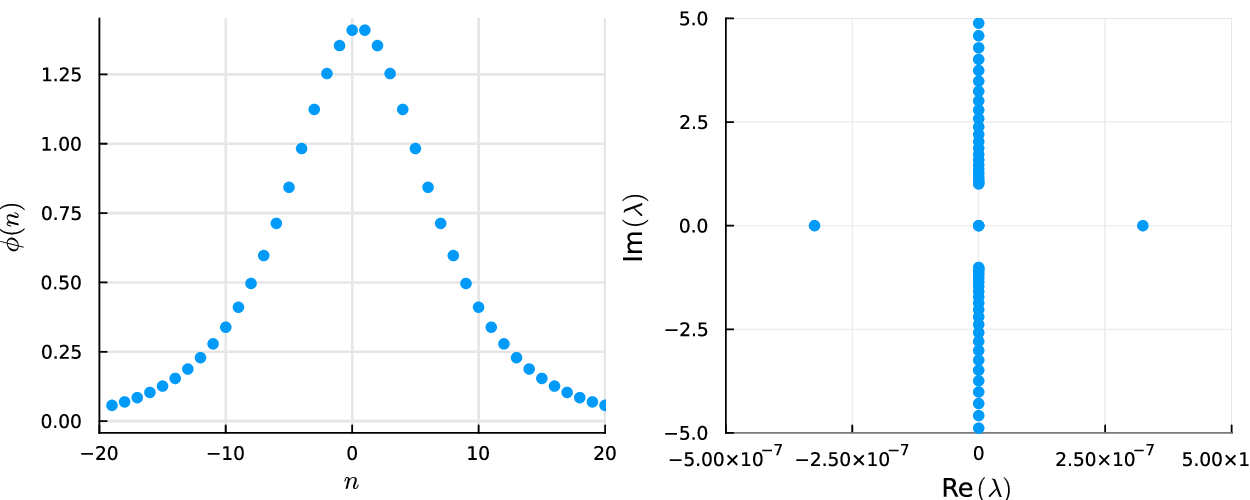}
		
		\caption{
			An intersite bright soliton (left) and its spectra in the complex plane (right) for $\varepsilon=0.2$. 
			The pair of real eigenvalues show that the soliton is unstable.}
		\label{fig:bright-sol-intersite}
	\end{figure*}
	
	\begin{figure*}
		\centering
		\includegraphics[width=0.8\linewidth]{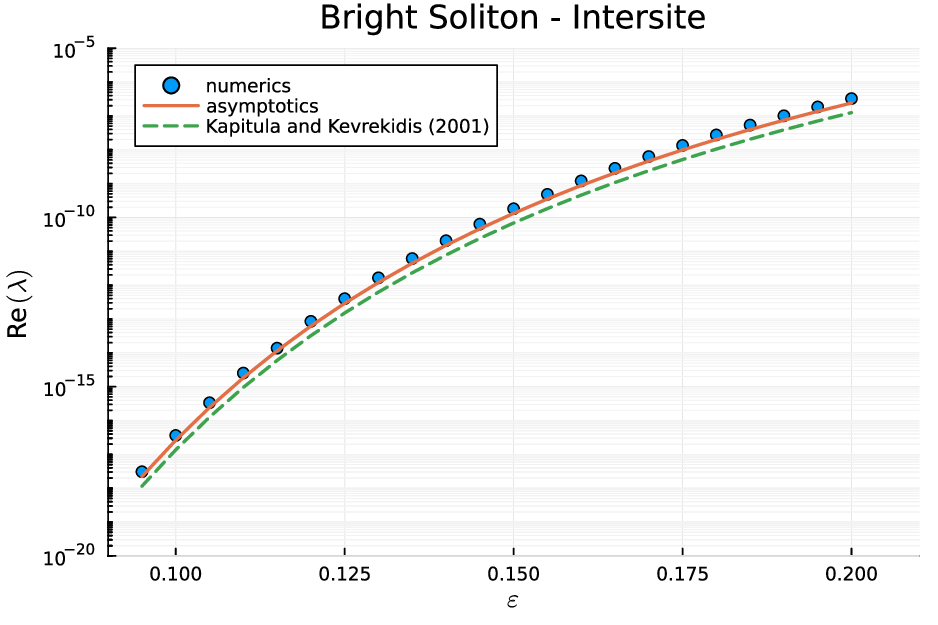}
		
		\caption{
			The unstable spectrum of intersite bright solitons as a function of the coupling $\varepsilon$. We plot our asymptotic result \eqref{eqn:dnls_bright_ev_approximation}, showing good agreement with the numerics. As a comparison, we also present the prediction by Kapitula and Kevrekidis \cite{Kapitula2001}. }
		\label{fig:bright-eigvals-intersite}
	\end{figure*}
	
	\begin{figure*}
		\centering
		\includegraphics[width=0.9\linewidth]{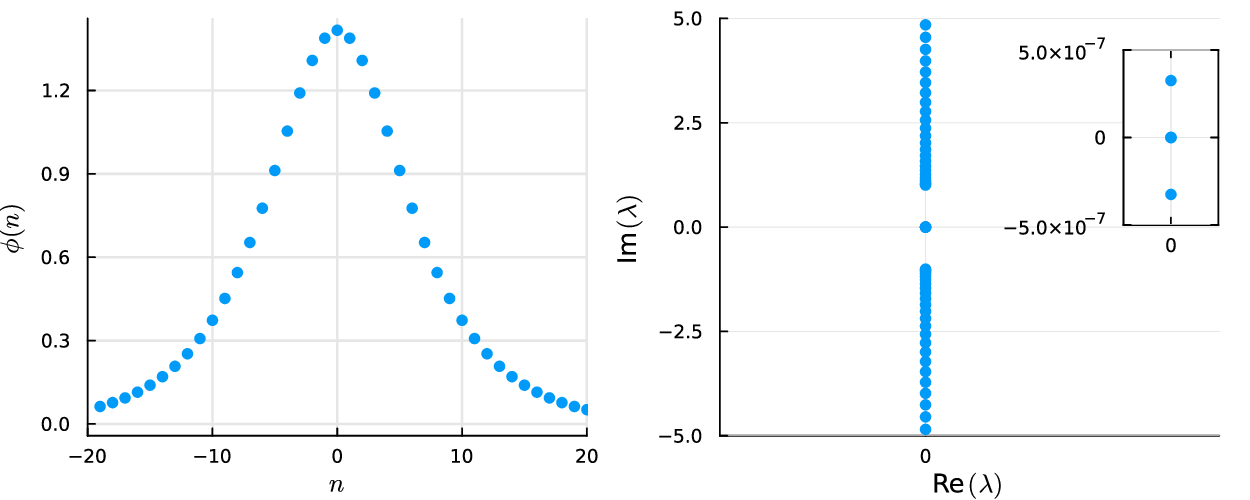}
		\caption{ Similar to Fig.\ \ref{fig:bright-sol-intersite}, but for onsite bright solitons. The inset on the left panel shows the presence of a pair of imaginary eigenvalues. }
		\label{fig:bright-sol-onsite}
	\end{figure*}
	
	\begin{figure*}
		\centering
		\includegraphics[width=0.8\linewidth]{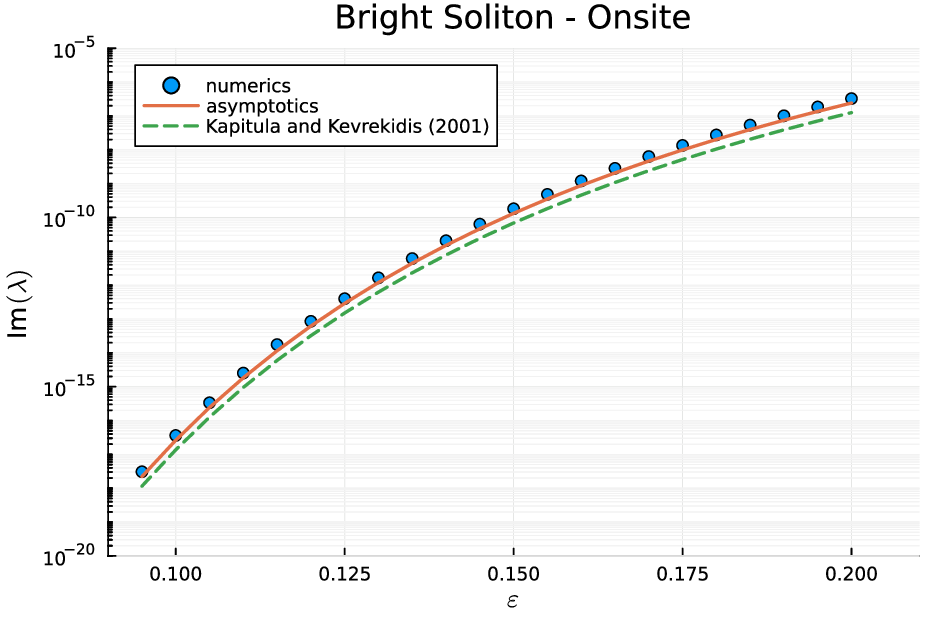}
		
		\caption{ Similar to Fig.\ \ref{fig:bright-eigvals-intersite}, but for onsite bright solitons. The numerical results show good agreement with our asymptotic prediction \eqref{eqn:dnls_bright_ev_approximation}. The result of Kapitula and Kevrekidis \cite{Kapitula2001} is also depicted.}
		\label{fig:bright-eigvals-onsite}
	\end{figure*}

	\section{Comparison with numerical results}
	\label{sec4}
	
	We solve the time-stationary DNLS equation numerically for both defocusing and focusing nonlinearities \eqref{eqn:dnls_main_cubic_defocusing_stationary} and \eqref{eqn:dnls_main_cubic_focusing_stationary}, together with their corresponding eigenvalue problems Eqs.\ \eqref{eqn:dnls_defocusing_evp} and \eqref{eqn:dnls_focusing_evp}, respectively.  We compute both the onsite and intersite solitons for each type of nonlinearity. We use arbitrary precision floating numbers BigFloat in Julia to capture small discrete eigenvalues.
	
	The computation of the stationary solutions is formulated as a system of nonlinear equations, with the initial guess $\phi^{(0)}(n) = \phi_0$ and $n_0 \in \{0,1/2\}$. Once a stationary solution $\tilde{\phi}$ is obtained, it will become the potential in the corresponding eigenvalue problem. We are interested in only a small number of eigenvalues to reduce the computational cost of solving the eigenvalue problem. We employ a shifted inverse power method \cite{ipsen1997computing}. 
	The standard iterative methods typically converge to the eigenvalue with the largest modulus. To serve our purpose, we instead apply the iteration to the matrix $\hat{L} = (\mathcal{L} - \delta I)^{-1}$, where $\mathcal{L}$ is the linear matrix from the eigenvalue problems in Eqs.\ \eqref{eqn:dnls_defocusing_evp} and \eqref{eqn:dnls_focusing_evp} and $\delta$ is a shifting parameter. We did not use the matrix $(L_+L_-)$, even though its size is smaller than using $\mathcal{L}$ because performing the iterations on the matrix will yield $\lambda^2$, which may compromise the available level of accuracy.

	By choosing an appropriate $\delta$, we transform the eigenvalue problem, allowing the iterative method to focus on the eigenvalues closest to $\delta$. Specifically, we set $\delta = 0$ for the intersite dark solitons and the onsite and intersite bright ones. The eigenvalue $\hat{\lambda}$ of $\hat{L}$ is related to the eigenvalue $\lambda$ of $\mathcal{L}$ by $\lambda = 1/\hat{\lambda}+\delta$. Eigenvalue calculations are performed using the Arnoldi method with Krylov-Schur restart \cite{Stewart2002KrylovSchur}, as implemented in ArnoldiMethod.jl. 
	
	We set the BigFloat precision to $2^{-400}\approx 10^{-121}$ in our numerical experiments. We consider a range of $\varepsilon$ values from 0.095 to 0.2 in increments of 0.05. Generally, small $\varepsilon$ values require large grid points 
	to capture the eigenvalues accurately. For our numerical calculations, we used $N_x =  600$ (intersite) or $N_x =  601$ (onsite)  grids for $\varepsilon \geq 0.015$ and $N_x=1200$ (intersite) or $N_x=1201$ (onsite) grids for smaller $\varepsilon$. The stationary solution is computed iteratively until the infinity norm of the residual is reduced to less than $10^{-50}$. For the dark soliton, an antisymmetric boundary condition was used, $\tilde{\phi}(-\lfloor N_x/2 \rfloor ) = -\tilde{\phi} (\lceil N_x/2 \rceil )$. For the bright soliton, the boundary condition is set as $\tilde{\phi}(-\lfloor N_x/2 \rfloor ) = \tilde{\phi} (\lceil N_x/2 \rceil ) =0$.

	A slightly different numerical approach is required to calculate 
	eigenvalues of the linear operator for the onsite dark soliton. As noted by Johansson and Kivshar \cite{PhysRevLett.82.85}, the soliton is unstable due to a pair of complex eigenvalues. A large number of grid points is required 
	to prevent discrete eigenvalues from entering the continuous spectrum. Thus, we start with computing eigenvalues with large $\varepsilon$ and small $N_x$ and calculate the discrete eigenvalues $\lambda(\varepsilon)$. Afterward, we apply the shifted inverse method to the eigenvalue problem $(\mathcal{L} - \delta I)^{-1}$. 
	We used previously computed eigenvalues as the shift parameter, i.e., $\delta = \lambda(\varepsilon)$. This ensures that the computation targets the eigenvalues of $\mathcal{L}$ closest to $\lambda(\varepsilon)$. If the computed eigenvalue $\lambda$ 
	is detected too close to the continuous spectrum (i.e., $|\mathrm{Re}\ (\lambda)|$ 
	is smaller than a threshold), it indicates that the eigenvalue is entering the continuous spectrum. In such cases, the computation is repeated with the number of grid points $N_x$ doubled, ensuring sufficient resolution to accurately resolve the real parts of the eigenvalues and preventing the entrance to the continuous spectrum. In this case, the main computational concern is the number of grid points $N_x$, whereas the numerical precision is less critical. Therefore, the calculations are performed with double precision. This iterative approach offers significant computational advantages: only one eigenvalue is computed per iteration, and the grid size is refined only when necessary, thus minimizing computational overhead while maintaining accuracy.
	
	An intersite dark soliton for \(\varepsilon = 0.2\) and its corresponding spectra are shown in Fig.\ \ref{fig:dark-sol-intersite}. The left panel presents the stationary solution, while the right panel shows the spectra in the complex plane. We observe a pair of exponentially small real eigenvalues, showing the instability of the intersite soliton. We also plot our asymptotics 
	given by Eq.\ \eqref{eqn:dnls_dark_ev_approximation}, where we obtain good agreement. 
	
	Figure \ref{fig:dark-eigvals-intersite} presents the numerically computed critical eigenvalues of the intersite dark soliton for varying the coupling constant $\varepsilon$. The asymptotic approximation demonstrates excellent agreement, showing the robustness of the asymptotic method in capturing the exponentially small eigenvalues of intersite dark solitons.
	
	For the onsite dark soliton, Fig.\ \ref{fig:dark-sol-onsite} presents the stationary solution (left panel) and the corresponding spectra (right panel). Unlike the intersite soliton, the onsite soliton exhibits a pair of complex eigenvalues, indicating oscillatory instability. Our analytical result cannot capture this instability because the asymptotic analysis shows that the soliton is supposed to have a pair of imaginary eigenvalues. However, our analytical calculation cannot consider that the imaginary axis is filled with the continuous spectrum. 
	
	Based on the expression of the asymptotic expansion (see Eqs.\ \eqref{eqn:dnls_dark_ev_approximation} and \eqref{eqn:dnls_bright_ev_approximation}), instead, we fit the calculated eigenvalue of the onsite dark soliton with the following ansatz: 
	\begin{equation} \label{eqn:dark_onsite_fitting}
		f(\varepsilon) =  c_1 \varepsilon^{-c_2} e^{-\frac{c_3}{\varepsilon}}. 
	\end{equation}
	The real and imaginary components of the eigenvalue are fitted separately using this expression to determine the respective parameter sets. For the real part, we use $c_1 = 8381.71$, $c_2 = -5.40$, and $c_3 = 13.79$, while for the imaginary part, $c_1 = 1237.69$, $c_2 = -4.69$, and $c_3 = 8.98$.
	
	We show the computational results for intersite bright solitons 
	in Fig.\ \ref{fig:bright-sol-intersite}. The right panel shows the soliton's spectra that consist of the continuous spectrum along the imaginary axis (with a gap in the interval \(-1i,1i\), zero eigenvalues of double multiplicity, and a pair of real eigenvalues indicating exponential instability of the soliton. We show the real eigenvalues of the intersite bright soliton in Fig.\ \ref{fig:bright-eigvals-intersite}, compared with the asymptotic approximation Eq.\ \eqref{eqn:dnls_bright_ev_approximation}. The numerical results demonstrate close agreement with the asymptotic prediction, even for relatively large 
	\(\varepsilon\). Additionally, we also provide a comparison with the asymptotic expansion obtained by Kapitula and Kevrekidis \cite{Kapitula2001}, i.e., 
	\begin{equation}
		\lambda^2 = -{(116.2)^2} \varepsilon^{-5} \exp\left(-\frac{\pi^2}{\varepsilon}\right) \cos(2\pi n_0),\label{kk}
	\end{equation}
	which is shown as a green dashed line. 
	
	In contrast, the onsite bright soliton, shown in Fig.\ \ref{fig:bright-sol-onsite}, has a conjugate pair of small imaginary eigenvalues, indicating oscillatory stability. These eigenvalues align with the theoretical prediction of stability for the onsite configuration. Figure \ref{fig:bright-eigvals-onsite} further illustrates the imaginary eigenvalues, showing excellent agreement between the numerical results and the asymptotic formula. We also compare it with the approximation \eqref{kk}.

	\section{Conclusion}\label{sec:conclusion}
	
	This study investigated the stability of discrete solitons in the DNLS equation with both focusing and defocusing nonlinearities. We derived analytical expressions for the eigenvalues that determine the linear stability of onsite and intersite solitons by employing exponential asymptotics. The results, confirmed by numerical simulations, provide a comprehensive understanding of these solitons' (in)stability characteristics, particularly for the instability of intersite dark solitons that was unavailable.
	
	Careful numerical computations demonstrate good agreement with the analytical predictions, where the eigenvalues are exponentially small. This agreement validates the accuracy of the exponential asymptotics approach and highlights its utility in resolving stability questions for discrete nonlinear systems. Overall, this study confirms prior results on the existence and stability of discrete bright solitons through an alternative methodology and demonstrates the versatility of exponential asymptotics for studying stability in lattice-based nonlinear systems.

	One promising direction for future work is to apply the exponential asymptotics framework developed in this study to examine the effects of additional nonlinear terms, such as competing nonlinearities \cite{petrov2015quantum,petrov2016ultradilute}, which model quantum droplets in Bose-Einstein condensates. These effects are experimentally relevant \cite{luo2021new,guo2021new,bottcher2020new}. Quantum droplets are self-bound states formed by a balance between attractive mean-field interactions and repulsive beyond-mean-field effects \cite{petrov2015quantum, cabrera2018quantum}. When confined in deep optical lattices \cite{morera2021universal,natale2022bloch,chomaz2022dipolar}, these droplets can become localized at discrete lattice sites, creating structures that are similar to discrete solitons described by the DNLS equation \cite{kusdiantara2024analysis}. The exponential asymptotic methods developed in our study will provide a useful framework for analyzing the existence and stability of such discrete solitons. The methods can also help analyze small radiation leakage that is important for understanding the long-time behavior of quantum droplets moving in optical lattices \cite{kartashov2024enhanced}, and which are not captured by standard asymptotic techniques.
    
    Another important direction of future work is extending the method to higher-dimensional DNLS systems \cite{malomed2002multidimensional,malomed2020nonlinearity}. This would allow for analyzing more complex soliton structures, such as vortices \cite{kevrekidis2015defocusing} and discrete Townes solitons \cite{bakkali2021realization}, whose stability properties remain less understood.

	After submitting this paper for publication, we learned that the same results for \emph{bright solitons} were reported independently using the same method in \cite{lustri2025exponential} (and \cite{lustri2025borelpade} that includes the next nearest neighbour coupling).

	\section*{Data availability}
	No data was used for the research described in the article.
	
	\section*{Declaration of competing interest} 
	The authors declare that they have no known competing financial interests or personal relationships that could have appeared to influence the work reported in this paper.
	
	\section*{CRediT authorship contribution statement}
	The manuscript was written with contributions from all authors. All authors have given their approval to the final version of the manuscript.\\
	
	\textbf{FTA}: Formal Analysis, Investigation, Software, Validation, Writing - Original Draft, Writing - Review \& Editing; 
	{\textbf{ANH}:Software, Writing – Original Draft, Writing – Review \& Editing;}
	{\textbf{RK}: Validation, Writing - Review \& Editing.}
	\textbf{HS}: Conceptualization, Methodology, Validation, Supervision, Writing - Original Draft, Writing - Review \& Editing.
	
	\section*{Declaration of generative AI and AI-assisted technologies in the writing process}
	
	During the preparation of this work, the authors used Grammarly and ChatGPT in order to improve language and readability. After using these tools/services, the authors reviewed and edited the content as needed and take full responsibility for the content of the publication.
	
	\section*{Acknowledgements}
	HS acknowledged support by Khalifa University through 
	a Competitive Internal Research Awards Grant (No.\ 8474000413/CIRA-2021-065) and Research \& Innovation Grants (No.\ 8474000617/RIG-S-2023-031 and No.\ 8474000789/RIG-S-2024-070). ANH acknowledges the contribution of Khalifa University's high-performance computing and research computing facilities in providing computational resources for this research. RK acknowledges Riset Utama PPMI FMIPA 2024 (617I/IT1.C02/KU/2024).
	
	\section*{Appendix: Equation for $S_{N}$ near the Stokes line}
	To analyze the behavior near the Stokes line, we define the inner angle variable $\tilde{\theta}$ as
	\begin{equation*}
		\tilde{\theta} = \dfrac{\theta - \left(-\frac{\pi}{2}\right)}{\delta(\varepsilon)}
	\end{equation*}
	where $\delta(\varepsilon) \ll 1$ and the order of $\delta$ is to be determined. 
	\par
	We first consider the exponential term on the right-hand side of \eqref{eqn:SN_eqn}, where we have
	\begin{align*}
		&(-1)^{N}e^{i\theta(2N-2+\beta)}e^{i\rho e^{i\theta}/\varepsilon}e^{-\rho/\varepsilon} \\
		=\ &\exp\Bigg[-i\pi N + i\theta(2N-2+\beta)+ie^{i\theta}\frac{\rho}{\varepsilon} - \frac{\rho}{\varepsilon}\Bigg] \\
		=\ &\exp\Bigg[-i\pi N + i\left(-\frac{\pi}{2}+\delta \tilde{\theta}\right)\left(2N - 2 + \beta\right) \\
		&\quad +i e^{i(-\pi/2 + \delta\tilde{\theta})}\frac{\rho}{\varepsilon} - \frac{\rho}{\varepsilon}\Bigg] \\
		=\ &\exp\Bigg[i\pi-i\frac{\pi}{2}\beta + i \delta\tilde{\theta}2N + i\delta\tilde{\theta}(\beta-2) + \frac{\rho}{\varepsilon}\left(e^{i\delta\tilde{\theta}}- 1\right)\Bigg] \\
		\sim\ &\exp\Bigg[i\pi-i\frac{\pi}{2}\beta + i\delta\tilde{\theta}\left(\frac{\rho}{\varepsilon} + 2\nu\right) + i\delta\tilde{\theta}(\beta-2) \\
		&\quad + \frac{\rho}{\varepsilon}\left(i\delta\tilde{\theta} - \frac{\delta^{2}\tilde{\theta}^{2}}{2} + O(\delta^{3})\right)\Bigg] \\
		\sim\ & \exp\Bigg[i\pi - i\frac{\pi}{2}\beta - \frac{\delta^{2}\tilde{\theta}^{2}}{2}\frac{\rho}{\varepsilon} + i\delta\tilde{\theta}(2\nu+\beta-2) + O\left(\frac{\delta^{3}}{\varepsilon}\right)\Bigg]
	\end{align*}
	We see that balance is given by the scaling $\delta(\varepsilon) = \sqrt{\varepsilon}$. With this, we may neglect the terms of order $O(\delta) = O(\sqrt{\varepsilon})$ and higher in the exponent, which leaves at leading order
	\begin{align*}
		&(-1)^{N}e^{i\theta(2N-2+\beta)}e^{i\rho e^{i\theta}/\varepsilon}e^{-\rho/\varepsilon} \sim -e^{-i\beta\pi/2} e^{-\rho\tilde{\theta}^{2}/2}
	\end{align*}
	Now, we can also calculate the remaining terms in the right-hand side $\left(\text{the term} \dfrac{\cosh(\kappa e^{-i\theta})-1}{e^{i2\theta} + 1}\right)$ at leading order to be
	\begin{align*}
		\frac{\cosh(\kappa e^{-i\theta}) - 1}{1 + e^{i2\theta}}
		&=\frac{\cosh(\kappa e^{-i(-\pi/2 + \delta\tilde{\theta})})-1}{1 + e^{i2(-\pi/2 + \delta\tilde{\theta})}} \\
		&=\frac{\cosh(i\kappa e^{i\delta\tilde{\theta}}) - 1}{1 - e^{i2\delta\tilde{\theta}}} \\
		&=\frac{\cos(\kappa e^{i\delta\tilde{\theta}})-1}{1 - e^{i2\delta\tilde{\theta}}} \\
		&= \dfrac{\cos\left[\kappa(1+i\delta\tilde{\theta}+O(\delta^{2}))\right] - 1}{1 - (1 + i2\delta\tilde{\theta} + O(\delta)^{2})}\\
		&\sim \frac{\cos\kappa + (\sin\kappa)(\kappa i \delta\tilde{\theta}) + \dfrac{(\cos\kappa)}{2!}\kappa^{2}\delta^{2}\tilde{\theta}^{2} - 1}{-i2\delta\tilde{\theta}} \\
		&\sim \dfrac{i\kappa^{2}\delta\tilde{\theta}}{4} \\
		&\sim \dfrac{i\kappa^{2}\sqrt{\varepsilon}\tilde{\theta}}{4}
	\end{align*}
	
	Thus, we obtain the equation of $S_{N}$ at leading order (as $\varepsilon \to 0$) in terms of the inner angle variable $\tilde{\theta}$ as \eqref{eqn:SN_eqn_inner_theta}
	\bibliographystyle{elsarticle-num}
	\bibliography{Report_dnls23Notes}
	
\end{document}